\documentstyle [epsf,12pt]{article}
\input amssym.def
\input amssym.tex
\begin{document}

\begin{titlepage}
\vspace*{-2cm}
\begin{flushright}CLNS 95/1373 \\ hep-th/9511073 
\end{flushright}
\begin{center}
\vspace{.4cm} {\Large {\bf Wilson Renormalization Group Study of
\vspace{.2cm}\\ Inverse Symmetry Breaking}}
\\ \vspace{1cm}
{\large Thomas G. Roos\footnote{Electronic address: roost@hepth.cornell.edu}}
\\ \vspace{.3cm}
{\em Newman Laboratory of Nuclear Studies, \\ Cornell University,
Ithaca, NY 14853} \\
\vspace{1.5cm}
\end{center}
\begin{abstract}
For a large class of field theories there exist portions of parameter
space for which the loop expansion predicts increased symmetry
breaking at high temperature. Even though this behavior would clearly
have far reaching implications for cosmology such theories have not
been fully investigated in the literature. This is at least partially
due to the counter intuitive nature of the result, which has led to
speculations that it is merely an artifact of perturbation theory. To
address this issue we study the simplest model displaying high
temperature symmetry breaking using a Wilson renormalization group
approach. We find that although the critical temperature is not
reliably estimated by the loop expansion the total volume of parameter
space which leads to the inverse phase structure is not significantly
different from the perturbative prediction. We also investigate the
temperature dependence of the coupling constants and find that they
run approximately according to their one-loop $\beta$-functions at
high temperature. Thus, in particular, the quartic coupling of
$\varphi^4$ theory is shown to increase with temperature, in contrast
to the behavior obtained in some previous studies.
\end{abstract}
\end{titlepage}

\section{Introduction}

Intuitively, one expects symmetries that are spontaneously broken at
$T=0$ to be restored at high temperature. Examples of this behavior in
condensed matter systems abound, and it was suggested many years ago
\cite{kirshnits+linde} that the same thing might happen in
relativistic field theories. This idea was quickly backed up by
calculations \cite{weinberg,dolan+jackiw}, which essentially showed
that for a general gauge theory with scalars and fermions the
quadratic part of the effective potential goes like some combination
of coupling constants times $T^2$. For theories with simple scalar
sectors this combination of couplings is quite generally
positive-definite, in which case the minimum of the effective
potential lies at the origin of field space for sufficiently high
temperatures. At the same time it was also pointed out
\cite{weinberg}, however, that for certain models the coefficient of
the $T^2$ term may be negative. In this case symmetries are not
restored at high temperature, and in fact theories which are symmetric
at $T=0$ may be broken for large $T$. While such theories are not
totally generic, they are by no means contrived: all that is necessary
is a sufficiently complicated scalar sector (two multiplets will often
suffice) and some constraints on the relative sizes of the couplings
in the theory. The later are usually weak enough so that the allowed
values for the couplings occupy a sizable portion of parameter space.
To be more precise: a necessary condition for high temperature
symmetry breaking is the existence of negative scalar couplings in the
Lagrangian. Such couplings, however, are automatically allowed in any model
with at least two scalar multiplets, and consequently in almost all
extensions of the standard model. The effect of these couplings on the
phase structure of a theory depends on the details of the model and
must be studied on a case by case basis.

It was quickly realized that theories with symmetry non-restoration or
multiple high temperature phase transitions have many attractive
phenomenological features. A partial list of applications includes 
the strong CP problem and GUT scale baryogenesis \cite{mohapat}, the
monopole problem \cite{sal+ska+stern,dvali}, baryogenesis and dark matter
\cite{brian}, and inflation \cite{lee+koh}.

In this paper we will not discuss any particular physical application
of models that exhibit an inverse phase structure. Instead, we will
focus on the phenomenon itself, specifically on the validity of the
one-loop calculation that predicts it in the first place. There are
several reasons for doing this, perhaps the most important being that
there seems to be a widespread suspicion that symmetry non-restoration
is merely an artifact of perturbation theory and not a true physical
effect. While this is partially due to the counter intuitive nature of
the phenomenon, there is some quantitative evidence to back up this
claim.  For example, a popular theory whose effective potential
predicts an inverse phase structure in the one-loop approximation is
the $O(N)\times O(N)$ model \cite{weinberg}. However, subsequent
calculations based on large~N expansions and Gaussian effective
potential techniques seemed to show that the symmetry is in fact
restored at high temperature \cite{restore}. While the validity of
these results is not clear to us, the fact that they incorporate some
non-perturbative physics clearly raises the specter that the conclusions
drawn from the loop expansion are erroneous.

There are additional reasons to be concerned about the validity of the
perturbative calculation of the effective potential. For example,
consider the case were the theory is symmetric at $T=0$, in which case
perturbation theory predicts that the symmetry will be broken at high
temperature.  It is well known that the loop-expansion breaks down in
the vicinity of the phase transition due to severe infrared
divergences. Now we expect these infrared effects to strongly
renormalize the coupling constants in the theory, which is important
because the existence of the symmetry breaking transition depends on a
certain relationship between these couplings. It is clearly not inconceivable
that the renormalization will destroy this relationship and circumvent
the transition all together. Even in the case where one starts in the
broken phase the situation is not entirely clear-cut. For while the
loop-expansion predicts that this symmetry will remain broken at
arbitrarily high temperature, it fails to include the effect of the
temperature on the coupling constants. Again it is possible that the effective
coupling constants appropriate at high temperature do not obey the 
inequalities required to keep the symmetry broken.

In light of all this it would be nice to study theories which naively
admit an inverse phase structure using techniques that: one, capture
enough non-perturbative physics to correctly handle the infrared
problems in the vicinity of second order (or weakly first order) phase
transitions; and two, unambiguously take into account the effects of
temperature on the parameters in the theory. Fortunately such a
technique has become available in recent years. It is based on an
exact renormalization group (RG) equation for the Wilsonian effective
action, which has the nice feature that it admits relatively simple
approximations that nonetheless capture a good deal of the
non-perturbative physics\footnote{See section \ref{sec:flow} for
details and references.}. Using this tool we will study a
${\Bbb Z}_2 \times {\Bbb Z}_2$ symmetric scalar field theory, the
simplest model that exhibits an inverse phase structure in the
one-loop approximation. As explained above, our goal will be to
test the validity of the perturbative calculation and to establish if
high temperature symmetry breaking is in fact possible.  

The paper is organized as follows. In section~\ref{sec:flow} we present a
brief derivation of the exact RG equation for the effective action and discuss
the approximate version that we will solve. To illustrate the techniques 
involved, section~\ref{sec:phi4} treats the high temperature phase 
transition of simple $\varphi^4$ theory. In section~\ref{sec:z2z2} we 
then focus on inverse symmetry breaking in the ${\Bbb Z}_2 \times {\Bbb Z}_2$
model. Section~\ref{sec:couplings} discusses the high temperature behavior
of coupling constants and its relation to phase transitions. Our results
are summarized in section~\ref{sec:concl}.

\section{Flow Equation}
\label{sec:flow}

For completeness we present a brief derivation of the exact
renormalization group equation. For details the reader is referred
to the literature \cite{morrisexact,morristrunc,wettexact}. 
We work in
D Euclidean dimensions and for simplicity consider a single real
scalar field $\varphi$.  For a theory with action $S[\varphi]$ a scale
dependent partition function is defined as (dots represent
contractions in function space)
\begin{equation}
\exp W_{\Lambda}[J] =\! N\!\!\!\int\!{\cal D}\varphi \exp
\{ -\frac{1}{2}\varphi 
\cdot\Delta^{-1}_{\Lambda}\! \label{eq:partfcn}
\cdot\varphi-S_{\Lambda_{0}}[\varphi]+J\!\cdot\!\varphi\},
\end{equation} 
where $\Lambda_{0}$ is an ultraviolet cutoff, $N$ is a $J$ independent
normalization factor, and
\begin{equation}
\Delta_{\Lambda} \equiv 
\frac{\theta_{\epsilon}(q,\Lambda)}{\theta_{\epsilon}(\Lambda,q)}\frac{1}{q^2} 
\end{equation}
is a free massless propagator 
times the ratio of two smooth 
(everywhere positive) cutoff functions $\theta_{\epsilon}$.
This function has the properties
\[ \theta_{\epsilon}(q,\Lambda)\approx 1 \;\; {\rm for} \;\; q >
\Lambda+\epsilon \] 
and
\[ \theta_{\epsilon}(q,\Lambda)\approx 0 \;\; {\rm for} \; \; q <
\Lambda-\epsilon \,.\] 
For later use we note that the sharp cutoff limit is given by
the Heaviside function:
\begin{equation}
\lim_{\epsilon \rightarrow
\,0}\theta_{\epsilon}(q,\Lambda)=\theta(q-\Lambda)\,. \label{eq:heaviside}
\end{equation}
The effect of the extra term in the exponential of
(\ref{eq:partfcn}) is to suppress the propagation of long wavelength
modes ($q^2 \ll \Lambda^2$) while leaving the ultraviolet modes
unaffected. In fact it is obvious from (\ref{eq:partfcn}) that

\begin{equation}
\lim_{\Lambda \rightarrow\, 0}W_{\Lambda}[J] = W[J]\;, \label{eq:limW}
\end{equation}

the usual generating functional of connected Green's functions.
The flow equation for $W_{\Lambda}$ is obtained by differentiating 
(\ref{eq:partfcn}):

\begin{eqnarray}
\frac{\partial W_{\Lambda}[J]}{\partial \Lambda} =
-\frac{1}{2}&&\left\{
\frac{\delta W_{\Lambda}} {\delta J}\cdot\frac{\partial
\Delta^{-1}_{\Lambda}}{\partial \Lambda} \cdot\frac{\delta
W_{\Lambda}} {\delta J} \right. \nonumber \\ && \left. + \;{\rm tr}
\left(\frac{\partial 
\Delta^{-1}_{\Lambda}}{\partial \Lambda}\frac{ \delta^{2}
W_{\Lambda}}{\delta J\,\delta J}\right)\right\}\,. \label{eq:rgw}
\end{eqnarray}

Now define a scale dependent Legendre effective action via
\begin{equation}
\Gamma_{\Lambda}[\phi]=-\frac{1}{2}\varphi
\cdot\Delta^{-1}_{\Lambda}\cdot\varphi - W_{\Lambda}[J] +
J\cdot\varphi\;, \label{eq:gamma}
\end{equation}
 where
\[ \varphi = \frac{\delta W_{\Lambda}}{\delta J}\;. \]
In terms of $\Gamma_{\Lambda}$, (\ref{eq:rgw}) becomes
\begin{equation}
\frac{\partial\Gamma_{\Lambda}[\varphi]}{\partial\Lambda} =
-\frac{1}{2}\, {\rm tr}
\left[\frac{1}{\Delta_{\Lambda}}\frac{\partial\Delta_{\Lambda}}{\partial
\Lambda}\left(1+\Delta_{\Lambda}\frac{\delta^{2}\Gamma_{\Lambda}}{\delta
\varphi\delta\varphi}\right)^{-1}\right]\,.\label{eq:dgamma}
\end{equation}
It is clear from (\ref{eq:limW}) and (\ref{eq:gamma}) that 
\begin{equation}
\lim_{\Lambda \rightarrow\, 0}\Gamma_{\Lambda}[\varphi] = \Gamma[\varphi]\;,
\label{eq:gam0}
\end{equation}
the generating functional of one particle irreducible (1PI) diagrams.
In the opposite limit, $\Lambda\rightarrow\infty$,
$\Delta^{-1}_{\Lambda}$ diverges, so that in (\ref{eq:partfcn}) the
``classical'' approximation to $W_{\Lambda}[J]$ becomes exact. Thus
\begin{equation}
W_{\Lambda}[J] \rightarrow -S[\varphi^{*}]-\frac{1}{2}\varphi^{*} 
\cdot\Delta^{-1}_{\Lambda}\cdot\varphi^{*}+J\cdot\varphi^{*}\;, 
\end{equation}
where \[ 
J=\left.\frac{\delta\,(S[\varphi]+
\frac{1}{2}\varphi \cdot\Delta^{-1}_{\Lambda}\cdot\varphi)}
{\delta\varphi}\right|_{\phi^{*}}\;.
\]
This says simply that as $\Lambda\rightarrow\infty$, $W_{\Lambda}$
becomes the Legendre transform of
$S+\frac{1}{2}\varphi\cdot\Delta^{-1}_{\Lambda}\cdot\varphi$. But from
(\ref{eq:gamma}) $\Gamma_{\Lambda} + \frac{1}{2}\varphi
\cdot\Delta^{-1}_{\Lambda}\cdot\varphi$ is the Legendre transform of
$W_{\Lambda}$, and since
$S+\frac{1}{2}\varphi\cdot\Delta^{-1}_{\Lambda}\cdot\varphi$ is always
convex in the limit $\Lambda\rightarrow\infty$, we obtain
\begin{equation}
\lim_{\Lambda\rightarrow\infty}\Gamma_{\Lambda}[\varphi] = S[\varphi]\;.
\label{eq:gaminf}
\end{equation}
We also have the approximate relation
\begin{equation}
\Gamma_{\Lambda_{0}}[\varphi] = S_{\Lambda_{0}}[\varphi]\;,\label{eq:bc}
\end{equation}
which follows from (\ref{eq:gaminf}) provided the UV cutoff
$\Lambda_{0}$ is sufficiently large compared to all mass scales in
$S_{\Lambda_{0}}$.  From (\ref{eq:gam0}) and (\ref{eq:gaminf}) we see
that $\Gamma_{\Lambda}$ interpolates between the classical action at
$\Lambda_{0}$ and the effective action at $\Lambda=0$. In fact,
$\Gamma_{\Lambda}$ has an interpretation as a Wilsonian quantum
effective action obtained by integrating out purely quantum modes with
momenta $q>\Lambda$ \cite{morrisexact}.


Equation (\ref{eq:dgamma}) is exact, but much too complicated to be solved
exactly. Its usefulness therefore hinges on the existence of sensible
approximation schemes. We begin by writing 
\begin{eqnarray}
\Gamma_{\Lambda} = \int d^{D}x [&&U_{\Lambda}(\varphi) + \frac{1}{2}
Z_{\Lambda}(\varphi)(\partial^{\mu}\varphi)^{2} \nonumber \\ 
&& + Y_{\Lambda}(\varphi,
(\partial^{\mu}\varphi)^{2})(\partial^{\mu}\varphi)^{4}]\;.
\end{eqnarray}
Plugging this into (\ref{eq:dgamma}), setting $\varphi$ equal to a
constant, and neglecting the $\varphi$ dependence of $Z_{\Lambda}$ we
arrive at an approximate evolution equation for the effective
potential:
\begin{eqnarray}
\frac{\partial U_{\Lambda}(\varphi)}{\partial\Lambda}=&&-\frac{1}{2}\int
\frac{d^{D}q}{(2\pi)^{D}}\,
\frac{\partial\theta_{\epsilon}(q,\Lambda)}{\partial\Lambda}\nonumber
\\ && \times\left\{\theta_{\epsilon}(q,\Lambda)
+\theta_{\epsilon}^{2}(q,\Lambda)[Z_{\Lambda}-1+
U^{\prime \prime}_{\Lambda} (\varphi)/q^{2}]\right\}^{-1} \label{eq:dUsmth}
\nonumber \\
\end{eqnarray}
(primes denote differentiation with respect to $\varphi$).  Taking the
sharp cutoff limit (\ref{eq:heaviside}) and dropping an infinite field
independent term we finally obtain\footnote{Taking the sharp cutoff
limit requires some care. We use the relation\cite{morrisexact} \[
\frac{\partial\theta_{\epsilon}(q,\Lambda)}{\partial\Lambda}
f(\theta_{\epsilon}(q,\Lambda), \Lambda)\;\rightarrow\;
-\delta(\Lambda-q)\int_{0}^{1}dt f(t,q)\;\;\; {\rm as}\;\;\;
\epsilon\rightarrow0
\]
(here $f(\theta_{\epsilon},\Lambda)$ must be continuous at $\Lambda=q$
in dependence on its second argument).}
\begin{equation}
\frac{\partial U_{\Lambda}(\varphi)}{\partial \Lambda}
=-\frac{K_{D}}{2}\Lambda^{D-1}\ln \left[Z_{\Lambda}+\frac{U^{\prime
\prime}_{\Lambda} (\varphi)}{\Lambda^{2}}\right]\;, \label{eq:dU}
\end{equation}
where $K_{D}=S_{D-1}/(2\pi)^{D}$, and $S_{D}$ is the surface area of a unit
D-sphere. 

This is the equation we will study in the present paper, but before
proceeding a few remarks are in order. First note that
(\ref{eq:dUsmth}) and (\ref{eq:dU}) have been obtained from the exact
result (\ref{eq:dgamma}) by a truncation of the operator basis. While
this is an uncontrolled approximation, the resulting equations have
been successfully applied to the study of phase transitions in two,
three, and four dimensions, at zero and finite temperature, including
the determination of accurate critical exponents
\cite{all,mark.wo.eps,wettphi4t,wett2scalt,wettcrit}. They thus
capture a large amount of the relevant physics. It is also possible to
show that (\ref{eq:dU}) is simply the first term in a systematic
expansion of the exact result\cite{morrisnew}, which provides some
theoretical justification for its remarkable success.  Finally, it is
interesting to note that if one fixes $U_{\Lambda}=U_{\Lambda_{0}}$
and $Z_{\Lambda}=1$ on the RHS of (\ref{eq:dU}), and then integrates,
the result is simply the well known one-loop effective potential. We
emphasize, however, that there are no loop corrections to
(\ref{eq:dU}): the use of $U_{\Lambda}$ on the RHS corresponds to a
resummation of an infinite subset of diagrams, to all orders in the
loop expansion.

In this paper we will study the phase structure of four dimensional
field theories at finite temperature. We will work with the sharp
cutoff equation (\ref{eq:dU}), which, for the purpose at hand, has at
least two advantages over its smooth cutoff counterpart: the flow
equations are simpler, and there is no dependence on the particular
choice of cutoff function. Using the sharp cutoff will also allow us
to compare the two approaches since finite temperature phase
transitions have been studied using a smooth cutoff in
\cite{wettphi4t,wett2scalt}.

\begin{figure}[htb]
\epsfxsize=5in
\centerline{\epsfbox{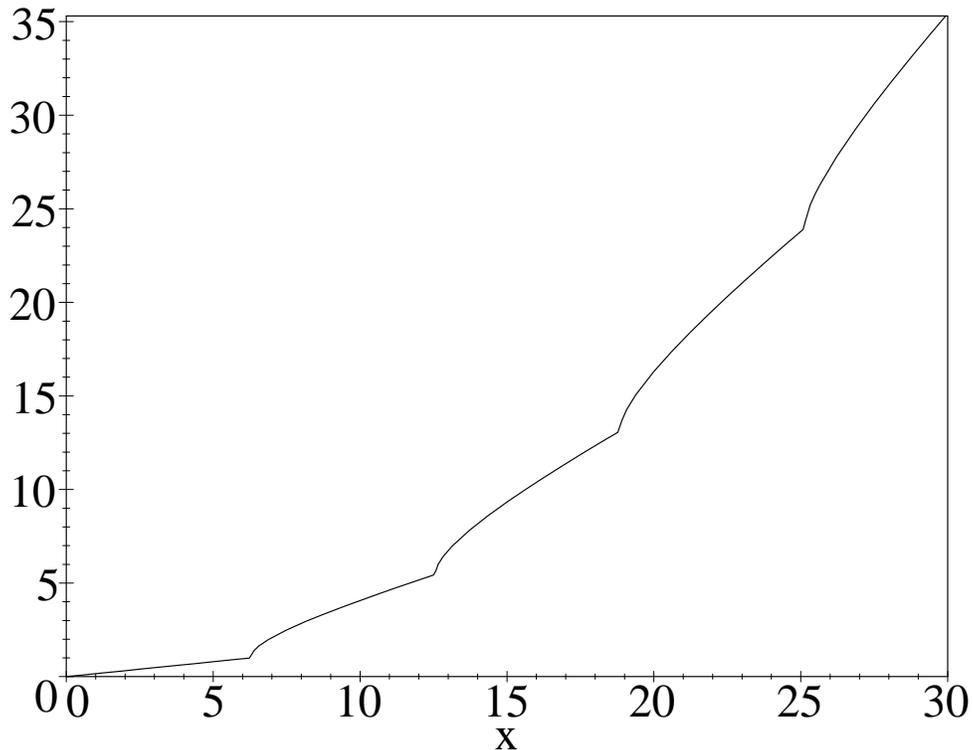}} 
\caption{The function g(x).}\label{fig:g}
\end{figure}

Specializing now to $D=4$ and setting $Z_{\Lambda}=1$
\footnote{A reasonable approximation for the theories at hand, since
the anomalous dimensions are known to be small
\cite{mark.wo.eps,wettcrit}.}, we are faced with the task of including
finite temperature effects.  We begin by writing (\ref{eq:dU}) as
\begin{equation}
\frac{\partial U_{\Lambda}(\varphi)}{\partial\Lambda}=-\frac{1}{2}\int
\frac{d^{4}q}{(2\pi)^{4}}\delta(\Lambda-q)\ln[q^{2}+U^{\prime\prime}_{\Lambda}
(\varphi)]\;. \label{eq:dUdelta}
\end{equation}
The imaginary time formalism instructs us to replace
\[
\int \frac{d^{4}q}{(2\pi)^{4}} \rightarrow T\sum_{n=-\infty}^{\infty}
\int \frac{d^{3}q}{(2\pi)^{3}}
\] and \[
q^2=q_{4}^{2}+{\bf q}^{2} \rightarrow \omega_{n}^{2}+ {\bf q}^{2}\;,\]
where $\omega_{n}=2\pi nT$.
This yields
\begin{equation}
 \frac{\partial
U_{\Lambda}(\varphi,T)}{\partial\Lambda}=-\frac{K_{3}}{2}2\pi
T^{2}\Lambda \,g\!\left(\frac{\Lambda}{T}\right)\ln[\Lambda^2 +
U^{\prime\prime}_{\Lambda} (\varphi,T)], \label{eq:dUT}
\end{equation}
where
\begin{equation}
g\left(\frac{\Lambda}{T}\right) \equiv
\left[\sum_{n=0}^{\left[\frac{\Lambda}{2\pi
T}\right]}2\sqrt{\left(\frac{\Lambda}{2\pi
T}\right)^{2}-n^{2}}\right]-\frac{\Lambda}{2\pi T}\;, \label{eq:g}
\end{equation}
and $[x]$ is the greatest integer $\leq x$. We note that $g(x)$ is
continuous and piecewise differentiable (see Fig.\ \ref{fig:g}).

It is easy to see that
\begin{equation}
g(x) \rightarrow \frac{x^{2}}{8\pi} \;\;\;\;\; {\rm as} \;\;\; x\rightarrow
\infty\;,
\end{equation}
so that (\ref{eq:dUT}) reduces to 
\begin{equation}
\frac{\partial
U_{\Lambda}(\varphi,T=0)}{\partial\Lambda}=-\frac{K_{4}}{2}\Lambda^{3}\ln
\left[\Lambda^{2}+U^{\prime \prime}_{\Lambda} (\varphi)\right]\;
\label{eq:dU4}
\end{equation}
in the limit $T \rightarrow 0$. Comparison with (\ref{eq:dU}) shows
that this is indeed the correct zero temperature equation. It is also
instructive to look at the opposite limit, $T \gg \Lambda$. Using
$g(x)=x/2\pi$ for $x < 2\pi$ we obtain
\begin{equation}
\frac{\partial U_{\Lambda}(\varphi,T)}{\partial
\Lambda}=-\frac{K_{3}}{2}T\Lambda^{2}\ln \left[\Lambda^{2}+U^{\prime
\prime}_{\Lambda} (\varphi)\right]\;\;\;\; {\rm for\;\;T\gg
\Lambda}. \label{eq:dUhighT}
\end{equation}
On comparing with (\ref{eq:dU}) we see that this is simply $T$ times the zero
temperature flow equation for a three dimensional theory. Hence we already
see evidence for dimensional reduction at high temperature.

For later use we record here the analog of (\ref{eq:dUT}) for theories
involving multiple scalar fields. The generalization is entirely 
straight forward, the result being
\begin{equation}
\frac{\partial U_{\Lambda}(\varphi,T)}{\partial \Lambda} =
-\frac{K_{3}}{2}2\pi T^{2}\Lambda
\,g\left(\frac{\Lambda}{T}\right)\ln\det\left[\Lambda^2 +
\frac{\partial^{2} U_{\Lambda}(\varphi)}{\partial
\varphi_{i}\partial\varphi_{j}}\right] \label{eq:dUTmult}
\end{equation}
where $g(\Lambda/T)$ is given by (\ref{eq:g}) as before.

Before concluding this section we point out that there exists in the
literature an alternative method of including finite temperature
effects in the sharp cutoff case \cite{liao1}. It amounts to
replacing $q$ by $|{\bf q}|$ in the delta function of
(\ref{eq:dUdelta}). The result is
\begin{equation}
 \frac{\partial U_{\Lambda}(\varphi,T)}{\partial \Lambda}\propto
\sum_{n=-\infty}^{\infty} \ln[\omega_{n}^{2}+\Lambda^{2}+
U^{\prime\prime}_{\Lambda} (\varphi,T)]\;. \label{eq:liao} 
\end{equation}
The sum over Matsubara frequencies may now be done analytically, and
the resulting equation involves no dependence on $n$, in contrast to
(\ref{eq:dUT}). However, by cutting off only the three momenta and
summing over all n one is neglecting the effect of integrating out the
Matsubara modes with $\omega_{n} > \Lambda$ on the effective
potential. As noted following (\ref{eq:dU}), it is precisely this kind
of feedback that leads to the incorporation of higher loop effects.
The equation derived from (\ref{eq:liao}) has the additional unpleasant feature
that it does not reduce to the zero temperature equation (\ref{eq:dU}) in the
limit $T\rightarrow 0$ \cite{liao2}.

\section{$\lambda\varphi^4$ Theory}
\label{sec:phi4}
In this section we will discuss briefly the finite
temperature phase transition for a simple ${\Bbb Z}_2$ symmetric scalar
field theory. This model has already been analyzed using the Wilson RG
approach \cite{wettphi4t,liao2}, and it is included here mainly to
illustrate the method. It is worth noting, however, that the previous
treatments differ from ours in the details of the implementation:
\cite{wettphi4t} uses a smooth cutoff, and in \cite{liao2} a sharp cutoff
is used only for the three-momentum, as discussed in the last paragraph of
the previous section. 

Consider then a ${\Bbb Z}_2$ symmetric theory with Lagrangian
\begin{equation}
{\cal L} = \frac{1}{2}\partial_{\mu}\varphi\partial^{\mu}\varphi-V(\rho)\;,
\end{equation}
where $\rho\equiv\frac{1}{2}\varphi^2$.
Our RG equation for the effective potential (\ref{eq:dUT}) becomes 
\begin{eqnarray}
 \frac{\partial U_{\Lambda}(\rho,T)}{\partial
\Lambda}=-\frac{\Lambda\,T^2}{2\pi}\,
g\left(\frac{\Lambda}{T}\right)\ln\left[ \Lambda^2 +
\frac{\partial U_{\Lambda}(\rho,T)}{\partial \rho}
+2\rho\frac{\partial^{2}U_{\Lambda}(\rho,T)}{\partial\rho^2}\right].
\label{eq:phi4} 
\end{eqnarray}
We wish to solve (\ref{eq:phi4}) subject to the boundary
condition\footnote{See (\ref{eq:bc}). Note that consistency requires that we
keep $T \ll \Lambda_{0}$ at all times.}
$U_{\Lambda_{0}}(\rho,T)=V(\rho)$.  This is still rather difficult,
and to make progress we expand $U_{\Lambda}$ in a Taylor series about
its minimum. This leads to an infinite set of coupled nonlinear
ordinary differential equations, which may be approximately solved by
truncation.  

Let us start with the symmetric regime, where the minimum
of $U_{\Lambda}$ is at the origin. We parameterize $U_{\Lambda}$ in terms
of its successive derivatives: 
\begin{equation}
m^{2}(\Lambda,T) \equiv U_{\Lambda}^{(1)}(\rho=0,T)\;,
\end{equation}
\begin{equation}
\lambda_{2n}(\Lambda,T) \equiv \frac{1}{n!}U_{\Lambda}^{(n)}(\rho=0,T)
\;\;\;{\rm for}\;\;\; n \geq 2 \;,
\label{eq:defl}
\end{equation}
where $U_{\Lambda}^{(n)}\equiv
\frac{\partial^{n}U_{\Lambda}}{\partial\rho^{n}}$.  Evolution
equations for these parameters are easily derived by differentiating
(\ref{eq:phi4}) with respect to $\rho$. The first three equations
obtained in this way are:
\begin{eqnarray}
\frac{dm^2}{d\Lambda} & = & -\frac{\Lambda\,T^2}{2\pi}\,
g\left(\frac{\Lambda}{T}\right)\frac
{6\,\lambda_{{4}}}{{\Lambda}^{2}+{m}^{2}}\; ,  
\label{eq:dm}
\\
\frac{d\lambda_{4}}{d\Lambda} & = & -\frac{\Lambda\,T^2}{2\pi}\,
g\left(\frac{\Lambda}{T}\right)
\left[\frac {15\,\lambda_{{6}}}{{\Lambda}^{2}+{m}^{2}}-\frac {18\,{\lambda_{{4}
}}^{2}}{\left ({\Lambda}^{2}+{m}^{2}\right )^{2}}\right], \label{eq:dl4}
\\
\frac{d\lambda_{6}}{d\Lambda} & = & -\frac{\Lambda\,T^2}{2\pi}\,
g\left(\frac{\Lambda}{T}\right)
\left[ 
{\frac {28\,\lambda_{{8}}}{{\Lambda}^{2}+{m}^{2}}}-{\frac {90\,\lambda_{{6}}
\lambda_{{4}}}{\left ({\Lambda}^{2}+{m}^{2}\right )^{2}}}
\right. \nonumber 
\\
&&~~~~~~~~~~~~~~~~~~~~~\left. +{\frac {72\,{
\lambda_{{4}}}^{3}}{\left ({\Lambda}^{2}+{m}^{2}\right )^{3}}}\right]\,.
\label{eq:dl6}
\end{eqnarray}

\begin{figure}[htb]
\epsfxsize=8cm
\centerline{\epsfbox{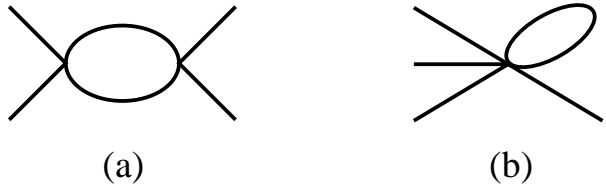}}
\caption{Contributions to the running of $\lambda_{4}$.}\label{fig:dl4}
\end{figure}

Several remarks are in order at this point. First, note that the above
expressions have an obvious interpretation in terms of one-loop
Feynman diagrams. For example, (\ref{eq:dl4}) says that the four point
function receives contributions from the diagrams in
Fig.\ \ref{fig:dl4}.
 Figure~\ref{fig:dl4}(a) corresponds to the second
term in square brackets of (\ref{eq:dl4}), which involves two 4-point
couplings and two ``propagators'', while Fig.\ \ref{fig:dl4}(b)
corresponds to the first term in square brackets, involving the
6-point coupling and one ``propagator''. The other flow equations can
be given similar interpretations.  We also see how integrating out the
high frequency modes generates new local interactions: even if one
starts with a $\varphi^{4}$ potential at $\Lambda_{0}$, higher order
terms are created as the cutoff is lowered.  Finally note that the
equations describe naturally how particles with mass larger than
$\Lambda$ ``decouple'', in the sense that their effect on the running
parameters becomes very small .

Now let us derive the flow equations in the broken regime, where the
minimum of the potential is at some $\rho_{0}>0$. Here we parameterize
$U_{\Lambda}$ in terms of the location of its minimum and consecutive
derivatives at that point. The flow equation for $\rho_{0}(\Lambda,T)$
is obtained from the condition
\begin{equation}
\frac{\partial U_{\Lambda}}{\partial \rho}(\rho_{0},T) = 0\;. \label{eq:uprime}
\end{equation}
Taking the $\Lambda$ derivative of (\ref{eq:uprime}) we obtain (primes
denote derivatives with respect to $\rho$):
\begin{equation}
\frac{\partial U^{\prime}_{\Lambda}}{\partial
\Lambda}+U^{\prime\prime}_{\Lambda} \,\frac{d\rho_{0}}{d\Lambda}=0\;,
\end{equation}
so that
\begin{equation}
\frac{d\rho_{0}}{d\Lambda}=-\frac{\partial U^{\prime}_{\Lambda}}{\partial
\Lambda}
\frac{1}{U^{\prime\prime}_{\Lambda}} \label{eq:drho1}
\end{equation}
(all $U_{\Lambda}$ derivatives are evaluated at $\rho_{0}$, of course).
The other parameters are defined as in (\ref{eq:defl}):
\begin{equation}
\lambda_{2n}(\Lambda,T) \equiv \frac{1}{n!}U_{\Lambda}^{(n)}(\rho_{0},T)\;,
\end{equation}
for $n \geq 2$.  The flow equations for these parameters are obtained
by differentiating (\ref{eq:phi4}) with respect to $\rho$ and evaluating the
result at $\rho_{0}$. The first three are:

\begin{equation}
\frac{d\rho_{0}}{d\Lambda} = \frac{\Lambda\,T^2}{2\pi}\,
g\left(\frac{\Lambda}{T}\right)\left[
{\frac {6\,\lambda_{{4}}+12\,\rho_{{0}}\lambda_{{6}}}{{\Lambda}^{2}+4\,\rho_
{{0}}\lambda_{{4}}}}\right]
\frac{1}{2\lambda_{4}}\; , \label{eq:drho}
\end{equation}
\begin{equation}
\frac{d\lambda_{4}}{d\Lambda} = -\frac{\Lambda\,T^2}{2\pi}\,
g\left(\frac{\Lambda}{T}\right)\left[
{\frac {15\,\lambda_{{6}}+24\,\rho_{{0}}\lambda_{{8}}}{{\Lambda}^{2}+
4\,\rho_{{0}}\lambda_{{4}}}}-{\frac {\left (6\,\lambda_{{4}}+12\,\rho_{{0
}}\lambda_{{6}}\right )^{2}}{2\,\left ({\Lambda}^{2}+4\,\rho_{{0}}\lambda_{{
4}}\right )^{2}}}\right]
+3\lambda_{6}\frac{d\rho_{0}}{d\Lambda}\; , \label{eq:dl4b}
\end{equation}
\begin{eqnarray}
\frac{d\lambda_{6}}{d\Lambda} & = & \! -\frac{\Lambda T^2}{2\pi}
g\!\left(\frac{\Lambda}{T}\right)\left[ 
{\frac {28\,\lambda_{{8}}+40\,\rho_{{0}}\lambda_{{10}}}{{\Lambda}^{2}+
4\,\rho_{{0}}\lambda_{{4}}}}-{\frac {\left (15\,\lambda_{{6}}+24\,
\rho_{{0}}\lambda_{{8}}\right )\left (6\,\lambda_{{4}}+12\,\rho_{{0}}
\lambda_{{6}}\right )}{\left ({\Lambda}^{2}+4\,\rho_{{0}}\lambda_{{4}}
\right )^{2}}} \nonumber \right. 
\\ 
& & \hspace{6.1em} +  \left. {\frac {\left (6\,\lambda_{{4}}+12\,\rho_{{0}}\lambda_{
{6}}\right )^{3}}{3\,\left ({\Lambda}^{2}+4\,\rho_{{0}}\lambda_{{4}}\right )
^{3}}}\right]
+4\lambda_{8}\frac{d\rho_{0}}{d\Lambda}. \quad\label{eq:dl6b} 
\end{eqnarray}

We begin our study of the high temperature phase transition by
computing the effective potential in the simplest possible
approximation. This amounts to parameterizing $U_{\Lambda}$ in terms of
its first two derivatives only, i.e., we set $\lambda_{n}=0$ for $n
\geq 6$. Our strategy is then as follows: we define the theory by
specifying ``renormalized'' zero temperature parameters at
$\Lambda=T=0$. We may choose the zero temperature theory to be in the
symmetric regime by picking $m^{2}_{r} \equiv m^{2}(0,0)>0$ or in the
broken regime by picking $\rho_{0r} \equiv \rho_{0}(0,0)>0$.  Depending
on the chosen phase, we then use either (\ref{eq:dm}),(\ref{eq:dl4}) 
or (\ref{eq:drho}),(\ref{eq:dl4b}) to numerically
integrate up in $\Lambda$ to some $\Lambda_{0} \gg m_{r} {\rm \;or\;}
\sqrt{\rho_{0r}}$. This ``integrating up'' is done at $T=0$,
and it serves merely to provide us with the ``bare'' parameters that
define the action $S_{\Lambda_{0}}$. We note that if $m^{2}$ or
$\rho_{0}$ go to zero at some intermediate $\tilde{\Lambda}$, then we
continue the integration with the set of equations appropriate to the
new phase, taking the values of the parameters at $\tilde{\Lambda}$ as
initial conditions.

\begin{figure}[htb]
\epsfxsize=5in
\centerline{\epsfbox{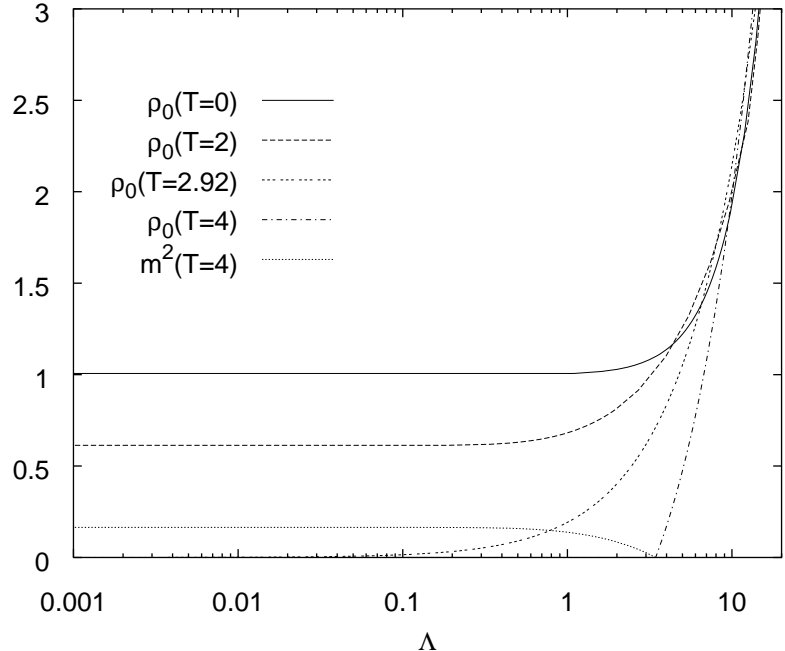}} 
\caption[The evolution of $\rho_{0}$ at various temperatures]{The
evolution of $\rho_{0}$ at various temperatures. For
$T>T_{c}$, the running of the mass term in the symmetric regime is
also shown.}\label{fig:rhovlam}
\end{figure}

\begin{figure}[htb]
\epsfxsize=5in
\centerline{\epsfbox{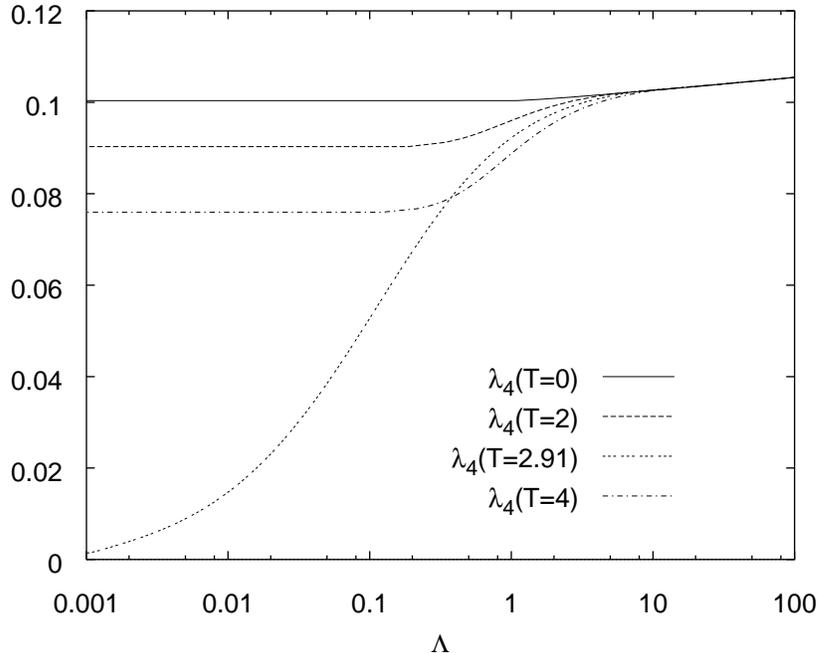}} 
\caption{The evolution of $\lambda_{4}$ at various
temperatures.}\label{fig:lam4vlam}
\end{figure}

With the parameters at $\Lambda_{0}$ in hand we are now ready to study
the effects of finite temperature. This is done by fixing $T \neq 0$
in the evolution equations and running down from $\Lambda_{0}$
to $\Lambda=0$. Repeating this for different values of T allows one to
determine the temperature dependence of the renormalized parameters.
Figures~\ref{fig:rhovlam} and~\ref{fig:lam4vlam} show the
evolution in $\Lambda$ of the parameters at various temperatures. The
initial conditions are such that the zero temperature theory is in the
broken phase.  From Fig.\ \ref{fig:rhovlam} we see that $\rho_{0}$
is quadratically renormalized for large $\Lambda$ and approaches a
constant as $\Lambda \rightarrow 0$. As the temperature is increased, the
asymptotic value of $\rho_{0}$ decreases, until at $T_{c}$ we have 
$\rho_{0}(\Lambda,T_{c})\rightarrow 0$ as $\Lambda \rightarrow 0$. Above
$T_{c}$ $\rho_{0}$ goes to zero for finite $\Lambda$, and we continue the
evolution in the symmetric regime.
Figure~\ref{fig:lam4vlam} shows the corresponding behavior of
$\lambda_{4}$.  At $T=0$ we observe the expected logarithmic
evolution. Above and below $T_{c}$, where all modes are massive, the
running of $\lambda_{4}$ is effectively stopped when $\Lambda$ becomes
much smaller than the relevant mass scale. At $T_{c}$, however,
$\lambda_{4}$ runs to zero. This is, in fact, expected, because at the
critical temperature the theory is scale invariant and the phase
transition is second order. As a consequence the parameters run
according to their canonical dimension, which is one for $\lambda_{4}$
at small $\Lambda$ since to the long wavelength modes the theory looks
effectively three dimensional at high temperature\footnote{This is
explained in more detail in the appendix.}. It is the fact that the
Wilson RG approach correctly takes into account the strong
renormalization of the coupling in the vicinity of the phase
transition that allows one to accurately describe the physics in this
region, including critical exponents
\cite{mark.wo.eps,wettphi4t,wettcrit}. This is to be contrasted with
the loop expansion, where higher order terms go like
$\lambda_{4}T/\cal{M}$ to some power, with $\cal{M}$ the appropriate
infrared cutoff. Since ${\cal M} \rightarrow 0$ at $T_{c}$, the
expansion necessarily breaks down near the phase transition.

\begin{figure}[htb]
\epsfxsize=5in
\centerline{\epsfbox{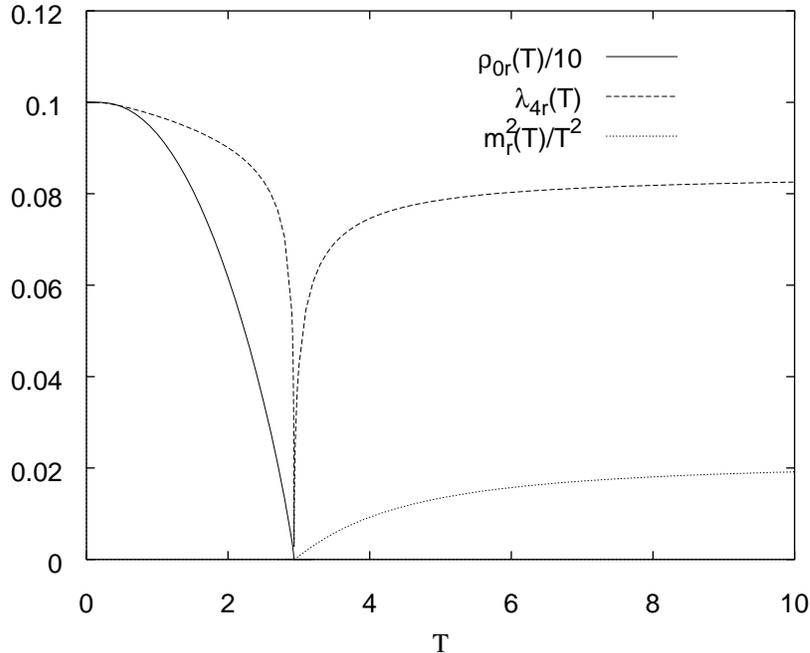}}  
\caption{The temperature dependence of the renormalized parameters.}
\label{fig:paramvt}
\end{figure}

Figure~\ref{fig:paramvt} shows the temperature dependence of the
``renormalized'' (i.e.\ $\Lambda=0$) parameters. The critical
temperature is given by $T^{2}_{c}/\rho_{0r} \approx 8.60$, whereas
naive one-loop perturbation theory predicts $T^{2}_{c}/\rho_{0r}=8$
\cite{weinberg,dolan+jackiw}. In light of the fact that the loop
expansion breaks down near the phase transition this good agreement
may seem surprising, but it is in fact expected: since perturbation theory
is valid both far above and far below\footnote{As long as there are no
Goldstone modes.} the critical temperature, one can show that the error
in the naive value of $T_{c}$ is of order $\lambda_{4}T_{c}$ \cite{weinberg}.
The above results may also be compared with \cite{wettphi4t}, where finite
temperature $\varphi^{4}$ theory was treated using a smooth cutoff flow
equation. 

Before going on to theories involving multiple scalar fields a
discussion of the validity of our approximation method is in order. We
will not comment on the initial step of obtaining (\ref{eq:dU}) from
(\ref{eq:dgamma}) by discarding the momentum dependence; for this we
refer the reader to the literature
\cite{morrisexact,morrisderiv,morrisnew}. What we do want to address
is the effect of truncating the infinite set of coupled differential
equations.  Instead of discussing the formal aspects of the problem
(see for example \cite{morristrunc,morrisnew}), we will focus on a
simple practical test: if the truncation is to be sensible at all,
then the effect of including additional terms must be, in some sense,
``small''.

To see if this is the case we derived flow equations analogous to
(\ref{eq:dm})--(\ref{eq:dl6}) and (\ref{eq:drho})--(\ref{eq:dl6b}) up
to and including $\lambda_{14}$. Solving this set of seven coupled
equations requires a change in tactics.  Recall that before we were
able to specify the renormalized ($\Lambda=0$) parameters and then
integrate up to obtain the ``bare'' parameters at $\Lambda_{0}$. This
is no longer possible, because the newly included couplings
$\lambda_{6}$--$\lambda_{14}$ correspond to irrelevant operators. This
means that their values in the infrared are fixed in terms of the
values of the relevant parameters $m^{2}$ and $\lambda_{4}$, and 
consequently they may not be chosen independently \cite{polch}. 

To isolate the effect of including the additional terms we therefore
proceed as follows. Start at $\Lambda_{0}$ and set
$\lambda_{6}$--$\lambda_{14}$ arbitrarily to zero\footnote{Any other
``natural'' value would do just as well, where ``natural'' means
${\cal O}(1)$ divided by enough powers of $\Lambda_{0}$ to get the
dimension right.}. Then fine tune $\rho_{0}(\Lambda_{0})$ and
$\lambda_{4}(\Lambda_{0})$ until flowing to the infrared produces
$\rho_{0r}=\rho_{0}(\Lambda=0)=1$ and
$\lambda_{4r}=\lambda_{4}(\Lambda=0)=0.1$. At this point we know we
are dealing with the same theory as before, and we may now switch
on the temperature and find $T_{c}$.  This procedure was carried out
for several different values of $\lambda_{4}$, the results being
summarized in table~\ref{tab:tc}. 
\begin{table}[tb]
\caption[$T_{c}^{2}/\rho_{0r}$ for various approximation methods and
couplings in the $\phi^4$ model]{$T_{c}^{2}/\rho_{0r}$ for various
approximation methods and
coupling values.}
\label{tab:tc}
\centerline{
\begin{tabular}{lccc}
\hline\hline & & $\lambda_{4r}$ & \\ \cline{2-4} Calculation Method~~~~~ &
~~~~~0.01~~~~~ & ~~~~~0.1~~~~~ & ~~~~~1.0~~~~~ \\ \hline  1-loop
Pert. Th. & 8 & 8 & 8 \\ Wilson RG up to $\lambda_{4}$ & 8.29 & 8.60 &
10.37 \\ Wilson RG up to $\lambda_{14}$ & 8.28 & 8.55 & ~9.69 \\
\hline\hline
\end{tabular}}
\end{table}
We see that at least for moderate
couplings the effect of including parameters beyond $\lambda_{4}$ is
small. This is in accord with the expectation that as long as the
effective potential is not too steep the first two terms in the Taylor
expansion should accurately describe its shape around the minimum.

\section{Inverse Symmetry Breaking}
\label{sec:z2z2}

The simplest model that exhibits an inverse phase structure is a
${\Bbb Z}_2 \times {\Bbb Z}_2$ symmetric scalar field theory described
by the Lagrangian
\begin{equation}
{\cal L} = \frac{1}{2}(\partial_{\mu}\varphi)^{2} +
\frac{1}{2}(\partial_{\mu}\chi)^{2} - V(\varphi,\chi) \; ,
\end{equation}
where
\begin{equation}
V(\varphi,\chi) = \frac{1}{2} m^2\varphi^{2} + \frac{1}{2}\mu^2\chi^{2}
+\frac{\lambda_{\varphi}}{4}\varphi^{4} +
\frac{\lambda_{\chi}}{4}\chi^{4} - \frac{\lambda_{\varphi\chi}}{2}
\varphi^{2}\chi^{2}.
\end{equation}
Note that, if the $\lambda$'s are all positive, we require 
\begin{equation}
\lambda_{\varphi}\lambda_{\chi} > \lambda_{\varphi\chi}^{2} \label{eq:bound}
\end{equation}
for boundedness.
If one calculates the finite temperature effective potential for this
theory to one loop and then expands the result about the origin one
finds that the quadratic part at high temperature is
\begin{eqnarray}
V_{\rm{1-loop}}^{\rm{quad}} &=& \frac{1}{2} \left(m^{2} +
\frac{T^{2}}{12}(3\lambda_{\varphi}
-\lambda_{\varphi\chi})\right)\varphi^{2} \nonumber \\ &+& \frac{1}{2}
\left(\mu^{2} + \frac{T^{2}}{12}(3\lambda_{\chi}
-\lambda_{\varphi\chi})\right)\chi^{2}\;.
\label{eq:1loop}
\end{eqnarray}
It is easy to see that this form allows a myriad of different symmetry
breaking patterns. Depending on the relative size of the couplings and
the signs of $m^2$ and $\mu^2$, one can have symmetry breaking, restoration,
or non-restoration at high temperature. Although less obvious, there
are even more exotic possibilities: for example, one can have a
symmetry broken or restored only for an intermediate range of temperatures
\cite{kephart}.

Let us now analyze this model using the Wilson RG formalism developed in the 
previous section. Because of the ${\Bbb Z}_2 \times {\Bbb Z}_2$
invariance the Lagrangian may be written as
\begin{equation}
{\cal L} = \frac{1}{2}(\partial_{\mu}\varphi)^{2} +
\frac{1}{2}(\partial_{\mu}\chi)^{2} - V(\rho,\zeta)\;,
\end{equation}
where
\begin{equation}
\rho \equiv \frac{1}{2}\varphi^{2} 
\end{equation}
and
\begin{equation}
\zeta \equiv \frac{1}{2}\chi^{2}\;.
\end{equation}
We proceed as in the previous section. The RG equation
(\ref{eq:dUTmult}) for this model is 
\begin{equation}
\frac{\partial U_{\Lambda}(\rho,\zeta,T)}{\partial\Lambda} =
-\frac{K_{3}}{2}2\pi T^{2}\Lambda
\,g\left(\frac{\Lambda}{T}\right)\ln\det\cal{M}\, , \label{eq:dUz2z2}
\end{equation}
where
\begin{equation}
\cal{M} = \left[
\begin{array}{cc} \Lambda^2 + 
\frac{\partial U_{\Lambda}}{\partial
\rho}+2\rho\frac{\partial^{2}U_{\Lambda} }{\partial\rho^2} &
2\rho^{\frac{1}{2}} \zeta^{\frac{1}{2}}
\frac{\partial^{2}U}{\partial\rho\partial\zeta} \\ 2\rho^{\frac{1}{2}}
\zeta^{\frac{1}{2}} \frac{\partial^{2}U}{\partial\rho\partial\zeta} &
\Lambda^2 + \frac{\partial U_{\Lambda}}{\partial
\zeta}+2\zeta\frac{\partial^{2}U_{\Lambda} }{\partial\zeta^2}
\end{array} \right]\,.
\end{equation}
The potential $U_{\Lambda}$ is again parameterized by its first
two derivatives at the minimum, but this time we must distinguish four
cases: the minimum may lie at the origin, on either axis, or between the 
axes. In all four cases the couplings are defined via
\begin{eqnarray}
\lambda_{\varphi}(\Lambda,T) & \equiv &
\left.\frac{1}{2}\frac{\partial^{2}U_{\Lambda}(T)}{\partial
\rho^{2}}\right|_{\rm min}\;,\\ \lambda_{\chi}(\Lambda,T) & \equiv &
\left.\frac{1}{2}\frac{\partial^{2}U_{\Lambda}(T)}{\partial\zeta^{2}}
\right|_{\rm min}\;,\\ \lambda_{\varphi\chi}(\Lambda,T) & \equiv &
\left.-\frac{1}{2}\frac{\partial^{2}U_{\Lambda}(T)}{\partial\rho
\partial\zeta}\right|_{\rm min}\;.
\end{eqnarray}
The other two parameters used to describe the potential depend on the
location of the minimum. If it is at the origin, we simply define
\begin{eqnarray}
m^{2}(\Lambda,T) & \equiv & \frac{\partial
U_{\Lambda}(0,0,T)}{\partial \rho}\;,\\
\mu^{2}(\Lambda,T) & \equiv & \frac{\partial
U_{\Lambda}(0,0,T)}{\partial \zeta}\;.
\end{eqnarray}
If the minimum is at $\rho_{0}$ on the $\rho$ axis, we use
\begin{equation}
\mu^{2}(\Lambda,T)  \equiv  \frac{\partial
U_{\Lambda}(\rho_{0},0,T)}{\partial \zeta}\;
\end{equation}
and $\rho_{0}$ to parameterize $U_{\Lambda}$. The flow equation for
$\rho_{0}$ is derived as in
(\ref{eq:uprime})--(\ref{eq:drho1}). Similarly, if the minimum is on
the $\zeta$ axis,
\begin{equation}
m^{2}(\Lambda,T)  \equiv  \frac{\partial
U_{\Lambda}(0,\zeta_{0},T)}{\partial \rho}\;
\end{equation}
and $\zeta_{0}$ are used. The final possibility is that the minimum
lies between the axes. In this case we parameterize in terms of its
location in field space, $(\rho_{0},\zeta_{0})$. The flow equations
for these coordinates are obtained by taking total $\Lambda$
derivatives of the two expressions
\begin{equation}
\frac{\partial U_{\Lambda}(\rho_{0},\zeta_{0})}{\partial \rho} = 0
\end{equation}
and
\begin{equation}
\frac{\partial U_{\Lambda}(\rho_{0},\zeta_{0})}{\partial \zeta} = 0\;.
\end{equation}
This yields two simultaneous algebraic equations for $d\rho_{0}/d\Lambda$
and $d\zeta_{0}/d\Lambda$, which may be solved to give
\begin{equation}
\frac{d\rho_{0}}{d\Lambda} =
-\frac{1}{2}\frac{\lambda_{\chi}\,\frac{\partial}{\partial\rho}\left.\left(
\frac{\partial U_{\Lambda}}{\partial\Lambda}\right)
\right|_{(\rho_{0},\zeta_{0})}
+\lambda_{\varphi\chi}\,\frac{\partial}{\partial\zeta}\left.\left(
\frac{\partial U_{\Lambda}}{\partial
\Lambda}\right)\right|_{(\rho_{0},\zeta_{0})}}{\lambda_{\varphi}\lambda_{\chi}-
\lambda_{\varphi\chi}^2} \;,
\end{equation}
\begin{equation}
 \frac{d\zeta_{0}}{d\Lambda} =
-\frac{1}{2}\frac{\lambda_{\varphi}\,\frac{\partial}{\partial\zeta}\left.\left(
\frac{\partial U_{\Lambda}}{\partial
\Lambda}\right)\right|_{(\rho_{0},\zeta_{0})}
+\lambda_{\varphi\chi}\,\frac{\partial}{\partial\rho}\left.\left(
\frac{\partial U_{\Lambda}}{\partial
\Lambda}\right)\right|_{(\rho_{0},\zeta_{0})}}{\lambda_{\varphi}\lambda_{\chi}-
\lambda_{\varphi\chi}^2} \;.
\end{equation}

As before, the right hand sides of the evolution equations are
expressed in terms of our parameters by differentiating
(\ref{eq:dUz2z2}) with respect to the fields, evaluating the result at
the appropriate minimum, and then dropping all terms with three or
more derivatives. For example, the flow equations for the symmetric
regime (i.e.\ minimum at the origin) are as follows:
\begin{eqnarray}
\frac{dm^2}{d\Lambda} & = & -\frac{\Lambda\,T^2}{2\pi}\,
g\!\left(\frac{\Lambda}{T}\right) \left[ \label{eq:dmz2z2}
\frac{6\,\lambda_{\varphi}}{\Lambda^2+m^2} 
-\frac{2\,\lambda_{\varphi\chi}}{\Lambda^2+\mu^2} \right], \\
\frac{d\mu^2}{d\Lambda} & = & -\frac{\Lambda\,T^2}{2\pi}\,
g\!\left(\frac{\Lambda}{T}\right) \left[
\frac{6\,\lambda_{\chi}}{\Lambda^2+\mu^2}
-\frac{2\,\lambda_{\varphi\chi}}{\Lambda^2+m^2} \right], \\
\frac{d\lambda_{\varphi}}{d\Lambda} & = & \frac{\Lambda\,T^2}{2\pi}\,
g\!\left(\frac{\Lambda}{T}\right) \left[
\frac{18\,\lambda_{\varphi}^2}{(\Lambda^2+m^2)^2}
+\frac{2\,\lambda_{\varphi\chi}^2}{(\Lambda^2+\mu^2)^2} \right], \\
\frac{d\lambda_{\chi}}{d\Lambda} & = & \frac{\Lambda\,T^2}{2\pi}\,
g\!\left(\frac{\Lambda}{T}\right) \left[
\frac{18\,\lambda_{\chi}^2}{(\Lambda^2+\mu^2)^2}
+\frac{2\,\lambda_{\varphi\chi}^2}{(\Lambda^2+m^2)^2} \right], \\
\!\frac{d\lambda_{\varphi\chi}}{d\Lambda} & = &
\frac{\Lambda\,T^2}{2\pi}\, g\!\left(\frac{\Lambda}{T}\right) \left[
\frac{6\,\lambda_{\varphi}\lambda_{\varphi\chi}}{(\Lambda^2+m^2)^2}
+\frac{6\,\lambda_{\chi}\lambda_{\varphi\chi}}{(\Lambda^2+\mu^2)^2}
\right.\nonumber \\ &&
~~~~~~~~~~~~~~~~~~\left.
-\frac{8\,\lambda_{\varphi\chi}^2}{(\Lambda^2+ m^2)(\Lambda^2+\mu^2)}
\right]. \label{eq:dlpxz2z2}
\end{eqnarray}
There are similar, albeit more complicated, sets of equations for
the other three positions of the minimum.

We begin our numerical study of the model by investigating the
phenomenon of inverse symmetry breaking. By this we mean choosing the
parameters so that the vacuum is symmetric at $T=0$ but asymmetric at
high temperature. From (\ref{eq:1loop}) it is clear that according to
the one-loop result this happens if, say,
$\lambda_{\varphi\chi}>3\lambda_{\varphi}$\footnote{ (\ref{eq:bound})
then requires that $\lambda_{\chi}>\lambda_{\varphi\chi}$, so the
other symmetry cannot be broken at high $T$.}. The critical temperature is
predicted to be
\begin{equation}
\frac{T_{c}^{2}}{m^{2}} = \frac{12}{\lambda_{\varphi\chi}-3\lambda_{\varphi}}
\;,\label{eq:tcnaive}
\end{equation}
independent of $\lambda_{\chi}$ and $\mu^{2}$.

The flow equations are solved as in the previous
section. Renormalized parameters are chosen at $T=\Lambda=0$ in such
a way that the vacuum is symmetric. The equations are then
integrated to some $\Lambda_{0}$ much larger than all physical
masses. Since the minimum moves away from the origin during this
process it was necessary to develop an algorithm that automatically
switches to the correct set of equations depending on the location of
the minimum. Once the parameters at $\Lambda_{0}$ are known the
temperature is switched on and the equations are integrated back
down to $\Lambda=0$, yielding the renormalized finite temperature values
for the masses, vacuum expectation values, and couplings. 

\begin{figure}[htb]
\epsfxsize=5in
\centerline{\epsfbox{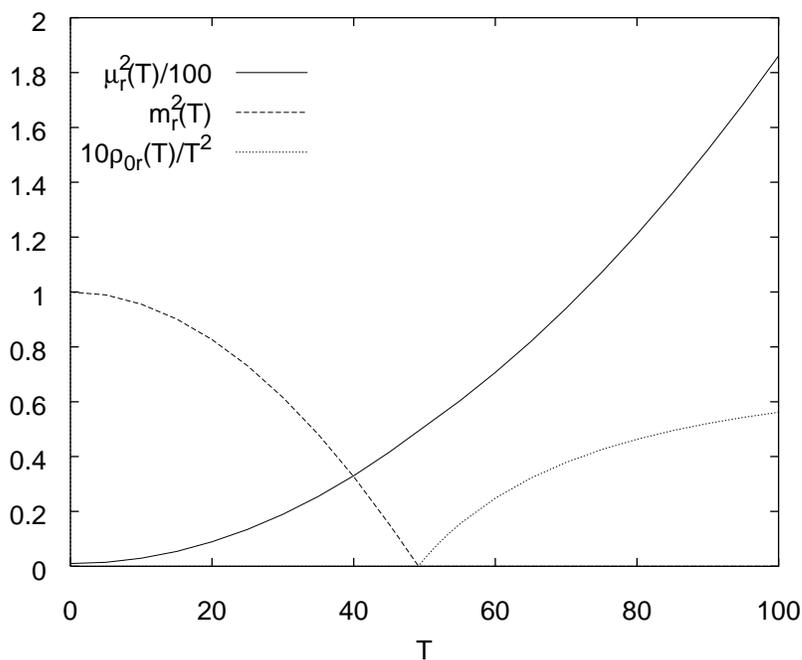}}   
\caption{The masses and $\rho_{0r}$ as a function of temperature.}
\label{fig:massesvt}
\end{figure}

\begin{figure}[htb]
\epsfxsize=5in
\centerline{\epsfbox{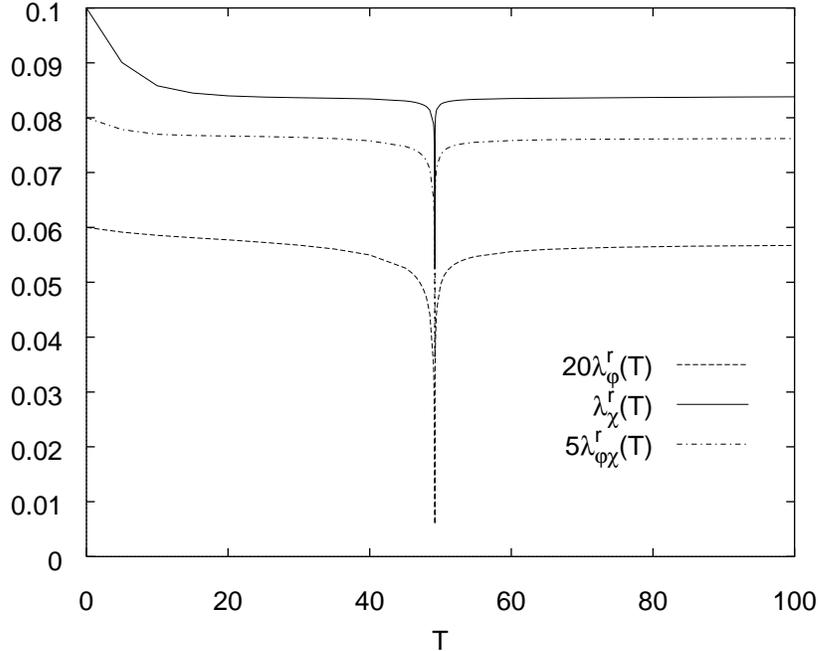}} 
\caption{The couplings as a function of temperature.}\label{fig:lambdasvt}
\end{figure}

Figures~\ref{fig:massesvt} and~\ref{fig:lambdasvt} show the behavior
of the renormalized parameters for a typical case. The parameters are
chosen so that naively (i.e.\ according to (\ref{eq:1loop})) the ${\Bbb Z}_2$
symmetry of the $\varphi$-field
is broken at high temperature. From Fig.~\ref{fig:massesvt} we see
that this picture is confirmed by our RG approach. As the temperature
increases $m^{2}_{r}(T)$ decreases and eventually hits zero at $T_{c}
\approx 49$. Both $\mu^{2}_{r}(T)$ and $\rho_{0r}(T)$ are proportional
to $T^2$ at high temperature. Figure~\ref{fig:lambdasvt} shows the
corresponding behavior of the couplings. As in the $\varphi^4$ case
discussed in the previous section, there is dramatic renormalization in
the vicinity of the phase transition\footnote{The behavior of the
couplings near $T_c$ is discussed in more detail in the appendix.}.

\begin{figure}[htb]
\epsfxsize=5in
\centerline{\epsfbox{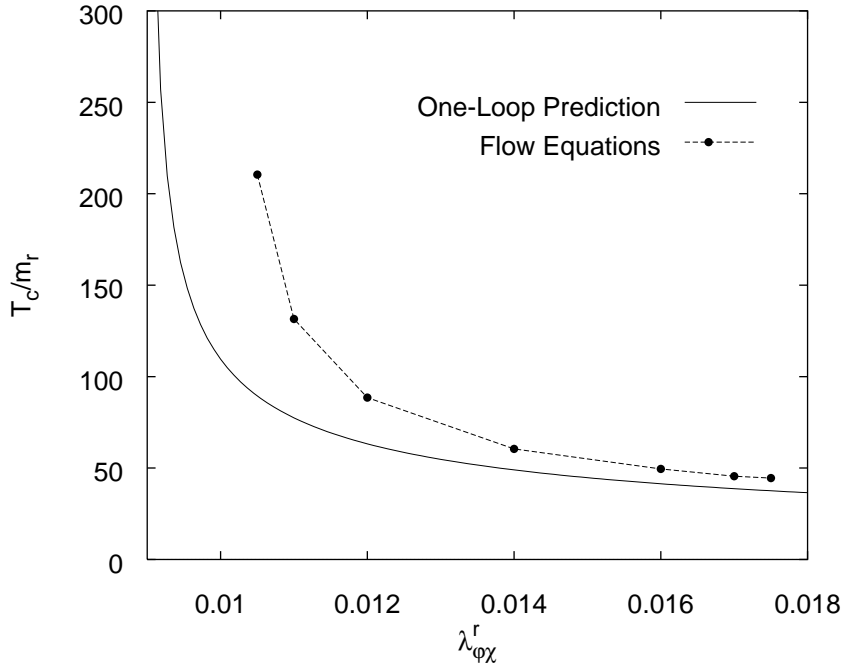}}   
\caption[Comparing the critical temperature obtained from one-loop
perturbation theory with that obtained by integrating the RG
equations]
{Comparing the critical temperature obtained from one-loop
perturbation theory with that obtained by integrating the RG equations.
($\lambda_{\varphi}^{r}=0.003$, $\lambda_{\chi}^{r}=0.1$, $\mu^2_r=1$)}
\label{fig:tcvalpha3}
\end{figure}

Figure~\ref{fig:tcvalpha3} compares the numerical value of the
critical temperature with the naive prediction, equation
(\ref{eq:tcnaive}), for several values of the cross-coupling. Even for
the relatively small couplings chosen there are significant
differences for all values of $\lambda_{\varphi\chi}^{r}$. The best
agreement occurs for large values of $\lambda_{\varphi\chi}^{r}$, but
even here the difference is about 15\%. (Note that (\ref{eq:bound})
requires that $\lambda_{\varphi\chi}^{r}<0.0173$. Numerically the
system becomes unstable slightly later, at $\lambda_{\varphi\chi}^{r}
\approx 0.018$.) As $\lambda_{\varphi\chi}^{r}$ decreases, the
deviation between the naive prediction and the numerical result
increases rapidly. For $\lambda_{\varphi\chi}^{r}=0.0105$, the
difference is more than a factor two. For
$\lambda_{\varphi\chi}^{r}=0.01$, perturbation theory predicts
$T_{c}/m_r \approx 109$. The numerical solution shows that there is no
phase transition at all for these values of the parameters, and that
$m^2_r(T)$ is in fact an increasing function of T at high temperature.

That the RG calculation gives a significantly higher critical
temperature than perturbation theory is due to a combination of two
factors. First of all, the renormalized couplings decrease with
temperature, which raises $T_{c}$. This effect, however, is also
present in the simple $\varphi^4$ case, and as was shown in section
\ref{sec:phi4}, there the two methods agree quite well. It alone
therefore cannot account for the difference in critical
temperatures. In fact, the dominant reason for the discrepancy is most
easily seen in Eq.\ (\ref{eq:dmz2z2}). The
``propagator'' in the second term is $\Lambda^2+\mu^2$, and $\mu^2$ is
a rapidly increasing function of temperature. Hence at high $T$ the
second term is suppressed compared to the first, which
significantly diminishes the effect of the cross-coupling
$\lambda_{\varphi\chi}$, and consequently raises $T_c$.

\begin{figure}[htb]
\epsfxsize=5in
\centerline{\epsfbox{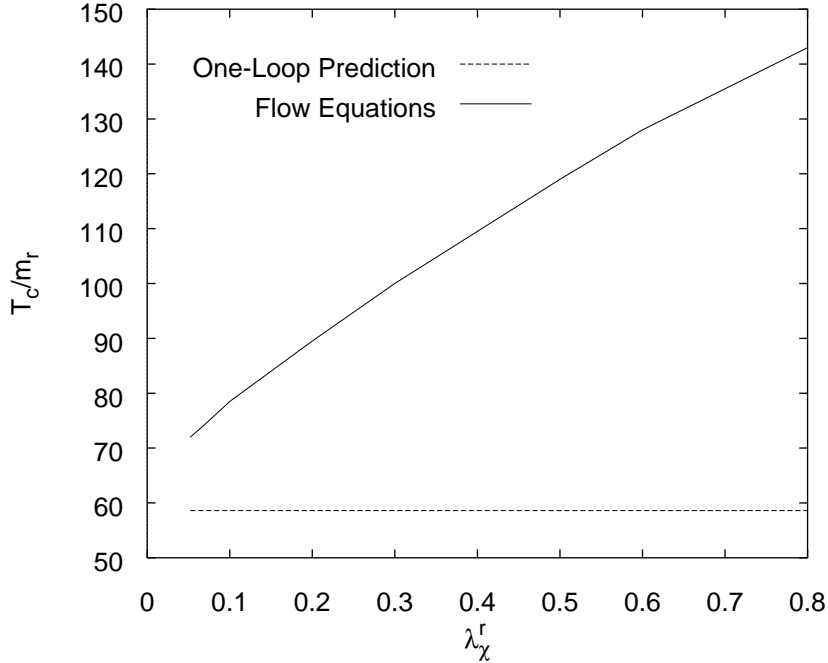}}  
\caption[The dependence of the critical temperature on
$\lambda_{\chi}^{r}$]{The dependence of the critical temperature on
$\lambda_{\chi}^{r}$.  ($\lambda_{\varphi}^{r}=0.003$,
$\lambda_{\varphi\chi}^{r}=0.0125$, $\mu^2_r=1$)}
\label{fig:tcvalpha2}
\end{figure}

We have just seen that the temperature dependence of $\mu^2$ plays an
important role in determining $T_c$. Since the leading correction to
$\mu^2$ is proportional to $\lambda_{\chi}$ we may therefore suspect
that the critical temperature also depends on the value of this
coupling, even though the one-loop result (\ref{eq:tcnaive}) predicts
that it does not. The question is settled by Fig.\ \ref{fig:tcvalpha2},
which shows the dependence of $T_c$ on $\lambda_{\chi}$. The effect is
clearly quite large and increases rapidly as the coupling
increases. Note that $\lambda_{\chi}>0.0521$ is required for
boundedness.

Figures~\ref{fig:tcvalpha3} and~\ref{fig:tcvalpha2} indicate that the
critical temperatures obtained from the RG approach are consistently
higher than those gotten from the loop expansion, and that certain
sets of couplings don't produce a phase transition at all even though
$\lambda_{\varphi\chi}> 3\lambda{\varphi}$. This may lead one to the
conclusion that the region of parameter space which yields an inverse
phase structure is significantly smaller than that predicted by the
one-loop result. We emphasize that this is not the case. The reason is
again the presence of the ``propagators'' in the flow equations. As
explained above, for the case at hand these factors reduce the
significance of the $\lambda_{\varphi\chi}$ term, which raises $T_{c}$
and suppresses the phase transition. But at the same time the suppression
of $\lambda_{\varphi\chi}$ tends to stabilize the system, which means
that there is additional parameter space beyond the limit (\ref{eq:bound})
for which transitions occur. The net effect is then a shift of the 
inverse-breaking region rather than a major reduction.
 
To make sure that the numerical results do not depend significantly on
the order of truncation of the evolution equations it is necessary to
study the effect of including higher dimensional operators. To this
end we enlarged our system of equations to include the couplings
corresponding to operators of dimension less than or equal to eight,
e.g., $\varphi^8$, $\varphi^2 \chi^4$, $\chi^6$, etc. This results in
a system of sixteen coupled first order nonlinear differential
equations. As discussed at the end of section~\ref{sec:phi4}, the
addition of irrelevant operators causes the complication that one can
no longer simply choose the renormalized parameters. Instead, one must
fine-tune the bare parameters at $\Lambda_{0}$ to achieve the desired
values for the masses and relevant couplings in the infrared. Aside
from this nuisance the solution of the equations proceeds as
before. The results are summarized in table~\ref{tab:tcz2z2}, which
shows that the effect of the added terms is very small.
\begin{table}[tb]
\caption[$T_{c}/m_{r}$ for various approximation methods and couplings
in the ${\Bbb Z}_2 \times {\Bbb Z}_2$ model]
{$T_{c}/m_{r}$ for various approximation methods and couplings
($\mu^2_r \approx 1$ in all cases). ``Up to dim. n'' means that
operators of dimension higher than n were discarded in the flow equations.}
\label{tab:tcz2z2}
\centerline{
\begin{tabular}{rccc}
\hline\hline $\lambda_{\varphi}^{r}=$ & $2.78\times 10^{-3}$ & $~~1 \times
 10^{-2}~~$ & $~~1 \times 10^{-4}~~$ 
\\ 
$\lambda_{\chi}^{r}=$ & $1.00 \times
 10^{-1}$ & $3\times 10^{-1}$ & $3\times 10^{-3}$ 
\\  
$\lambda_{\varphi\chi}^{r}=$ & $1.25\times 10^{-2}$ & $5\times
 10^{-2}$ & $5\times 10^{-4}$ 
\\ 
\hline 1-loop Pert. Theory~~ & 53.7 & 24.5 & 245 
\\ 
RG up to dim. 4 ops. & 67.9 & 33.6 & 257 
\\ 
RG up to dim. 8 ops. & 67.9 & 33.5 & 257 
\\ \hline\hline
\end{tabular}}
\end{table}
We conclude that discarding operators of mass dimension higher than four
is a valid approximation for the purpose at hand. Table~\ref{tab:tcz2z2}
also shows that for very small values of the couplings the numerical
result is in reasonable agreement with the one-loop prediction, at least
if $\lambda_{\varphi\chi}$ is safely larger than $3\lambda_{\varphi}$. For the
couplings in column~2 the two methods differ by almost 40\%, while the
discrepancy is down to 5\% for the values in column three.

\section{The Coupling Constants at High Temperature}
\label{sec:couplings}

In this section we discuss the temperature dependence of the coupling
constants in the high T limit. We begin with the simplest case, that
of a single scalar field.

\subsection{$\lambda\varphi^4$ Theory}

The high temperature behavior of the quartic coupling
$\lambda$ has been the subject of controversy for some time. Several
authors have found $\lambda$ decreasing with $T$
\cite{josephfujiliao2}, others find $\lambda$ approaching a 
constant as $T\rightarrow\infty$ \cite{chia}, and still others find
that $\lambda$ is increasing at high temperature \cite{weldon}. 

\begin{figure}[htb]
\epsfxsize=5in
\centerline{\epsfbox{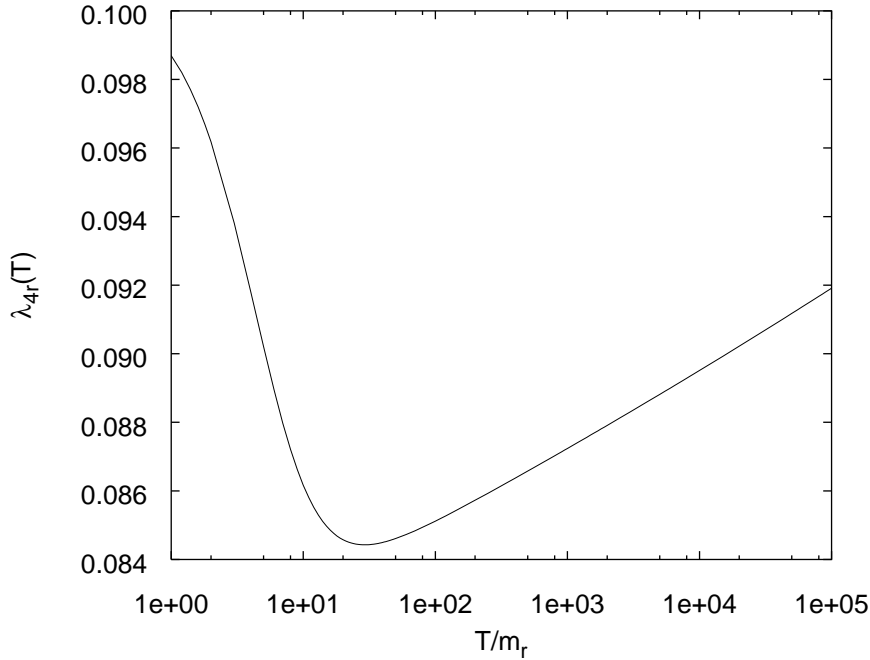}}   
\caption[Temperature dependence of the quartic coupling constant]
{Temperature dependence of the quartic coupling constant. Note the
logarithmic scale on the temperature axis. ($\lambda_{4r}=0.1$)}
\label{fig:lam4vbigt}
\end{figure}

In order to shed some light on the issue we integrated the RG
equations discussed in section~\ref{sec:phi4} for a wide range of
temperatures.  The results are shown in Fig.\ \ref{fig:lam4vbigt}. We
see that the coupling constant rapidly decreases at first, but
eventually turns around. From $T\approx 40$ onwards the dependence is
clearly approximately logarithmic. We note that at least qualitatively
this behavior agrees with that obtained in \cite{fendley}.  
 
\begin{figure}[htb]
\epsfxsize=5in
\centerline{\epsfbox{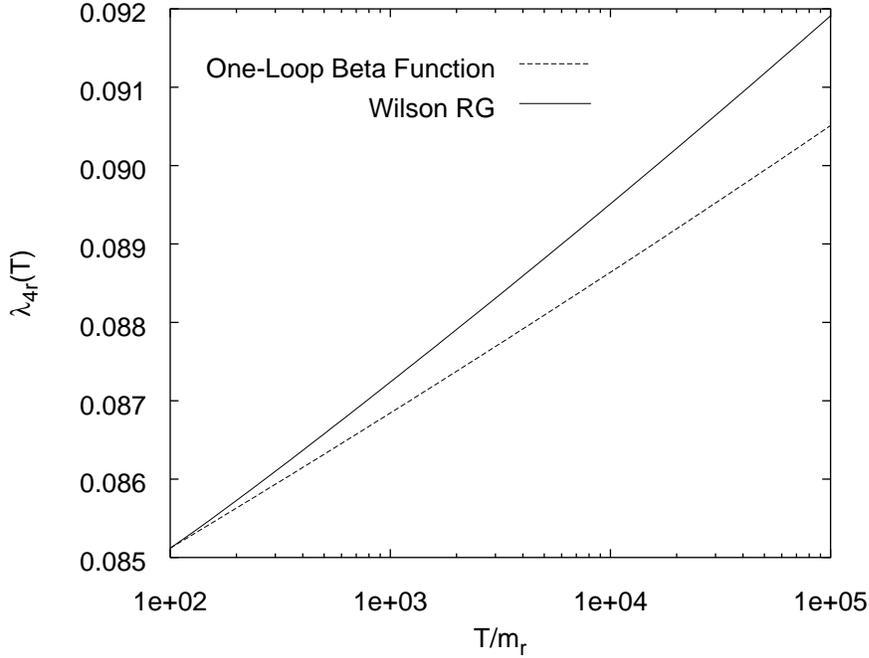}} 
\caption[The high temperature evolution of $\lambda_{4r}$(T) obtained 
from the RG compared to the behavior obtained by using $T$ as the scale
in the usual one-loop $\beta$-function]
{The high temperature evolution of $\lambda_{4r}$(T) obtained 
from the RG compared to the behavior obtained by using $T$ as the scale
in the usual one-loop $\beta$-function. The initial condition for the
one-loop integration is chosen so that the two methods agree at
T=100. ($\lambda_{4r}=0.1$)}
\label{fig:1loopbeta}
\end{figure}

In the literature one frequently encounters the notion that the effect
of high temperature on coupling constants may be incorporated by
``running'' the couplings according to their one-loop $\beta$-functions
using the temperature as the scale \cite{tscale}. In our case this
amounts to assuming
\begin{equation}
\frac{d\lambda_{4r}}{dt} = \frac{9}{8\pi^2}\lambda_{4r}^2\;, \label{eq:beta} 
\end{equation}
where $t=\ln (T/T_{0})$, so that
\begin{equation}
\lambda_{4r}(T) =\frac{\lambda_{4r}(T_{0})}{1-\frac{9}{8\pi^2}
\lambda_{4r}(T_{0})\ln (T/T_{0})}\;. \label{eq:1looplam}
\end{equation}
The justification for this procedure is usually based on rather vague
arguments concerning the average momentum transfer during collisions
of particles in the heat bath being ${\cal O}(T)$, so that a coupling
constant at that scale ought to be appropriate. While this is
certainly not unreasonable it is hardly a convincing argument, and as
noted above a more detailed analysis has produced many other types of
behavior. To see how well Eq.\ (\ref{eq:beta}) describes the high
temperature evolution we have plotted Eq.\ (\ref{eq:1looplam})
together with the numerical integration in
Fig.\ \ref{fig:1loopbeta}.  The one-loop result is normalized so
that the two methods agree at $T/m_{r}=100$.
It is clear from the figure that the high temperature dependence of
$\lambda_{4r}(T)$ is described quite well by the one-loop 
$\beta$-function.  After running $T$ over three orders of magnitude the
results differ by about 1.5\%. It is also clear, however, that the
difference becomes significant if one runs over many orders of
magnitude, for example to the GUT scale. In addition we see that using
the one-loop equation is only reasonable if one knows the correct
``initial condition'' at high temperature, in our case
$\lambda_{4r}(T)\approx 0.085$ at $T/m_{r}=100$. Starting the
evolution using Eq.\ (\ref{eq:1looplam}) with $\lambda_{4r}(T=1)=0.1$
is obviously not satisfactory, and obtaining the proper initial
condition requires an analysis such as the one presented in this paper.

It is worthwhile noting that the general features of the high
temperature behavior of the coupling can be understood by simply
looking at its RG equation, obtained in section~\ref{sec:phi4}. For
$\lambda_{6}=0$ and $\Lambda \gg T$ and $|m|$, Eq.\ (\ref{eq:dl4})
reduces to Eq.\ (\ref{eq:beta}). Thus the evolution for very large
$\Lambda$ is simply given by the one-loop $\beta$-function. Now for $T
\gg m_r$ the propagator in (\ref{eq:dl4}) contains a large thermal
mass, which means that the running effectively stops for $\Lambda <
\sqrt{\lambda_{4}} T$. The evolution for $\Lambda\ll
T$ is thus strongly suppressed, and all that remains is the
one-loop-like running for large $\Lambda$ plus some threshold
effect for $\Lambda \sim T$.  At a higher temperature the thermal mass
is larger, causing the running of $\lambda_{4}$ to freeze out at larger
$\Lambda$. The contribution from the threshold region is approximately
the same as before, and so the net effect of increasing $T$ from $T_1$
to $T_2$ is essentially equivalent to integrating the
one-loop $\beta$-function between the two temperatures. This explains
the correspondence in Fig.\ \ref{fig:1loopbeta}. The slope of the
numerical curve is slightly steeper because the quantum corrections to
the mass go like $-\lambda_{4}\Lambda^2$, which reduces the
denominator in (\ref{eq:dl4}) and makes $\lambda_{4}$ flow somewhat
faster for large $\Lambda$ than predicted by Eq.\ (\ref{eq:beta}).

Having just explained why $\lambda_{4r}(T)$ should increase with
temperature we find ourselves having to account for the initial
decrease seen in Fig.\ \ref{fig:lam4vbigt}. This decrease occurs for
values of the temperature low enough so that the suppression of the
evolution due to the thermal mass is not yet important. Instead the
running is dominated by the range $\Lambda < T$, where the prefactor
on the RHS of Eq.\ (\ref{eq:dl4}), $\Lambda T^2 g(\Lambda/T)$, is
a rapidly increasing function of $T$. Thus, for moderate values of
$T$, the coupling runs faster at higher temperature, which explains
the initial decrease in $\lambda_{4r}(T)$.

We end this section with a few remarks. First of all, we have checked
that the high temperature behavior of the coupling is not altered by
the inclusion of higher dimensional operators in the flow equations. In
fact, the effect of these operators decreases rapidly as one gets
further away from the critical theory. Second, we point out that the
author of \cite{fendley}, using an ``improved'' loop expansion
approach, has found the high temperature evolution of the coupling to
be similar to ours, i.e., approximately as predicted by the one-loop
$\beta$-function with some temperature dependent correction.  Finally,
note that the calculations in this section were done for a theory that
is symmetric at $T=0$. This choice has no effect on the high
temperature behavior of the coupling, for at very high T the theory
does not care whether or not there was a phase transition at much
lower temperature\footnote{This was also checked numerically.}. For the
temperature dependence of the coupling constant in an initially broken
theory at moderate values of T see section~\ref{sec:phi4}.

\subsection{${\Bbb Z}_2 \times {\Bbb Z}_2$ Model}

We will now study the high temperature behavior of the coupling
constants of the ${\Bbb Z}_2 \times {\Bbb Z}_2$ model discussed in
section~\ref{sec:z2z2}. There are two possible phases at high
temperature: either the theory is completely symmetric, or one of the two
symmetries is broken. We have checked that in both cases for generic
values of the zero temperature parameters the evolution of the
couplings at high temperature is approximately given by their one-loop
$\beta$-functions, viz.,
\begin{eqnarray}
\frac{d\lambda_{\varphi}}{dt} & = & \frac{1}{8\pi^2}[9\lambda_{\varphi}^2
+\lambda_{\varphi\chi}^2]\; , \label{eq:beta1} \\
\frac{d\lambda_{\chi}}{dt} & = & \frac{1}{8\pi^2}[9\lambda_{\chi}^2
+\lambda_{\varphi\chi}^2]\; , \\
\frac{d\lambda_{\varphi\chi}}{dt} & = & \frac{1}{8\pi^2}[3(\lambda_{\varphi}+
\lambda_{\chi})\lambda_{\varphi\chi}-4\lambda_{\varphi\chi}^2]\; ,
\label{eq:beta3}
\end{eqnarray}
where $t = \ln (T/T_0)$. The reason for this behavior is essentially
the same as in the $\varphi^4$ case discussed above, except that the 
situation is complicated by the presence of multiple mass scales.
\begin{figure}[htb]
\epsfxsize=5in
\centerline{\epsfbox{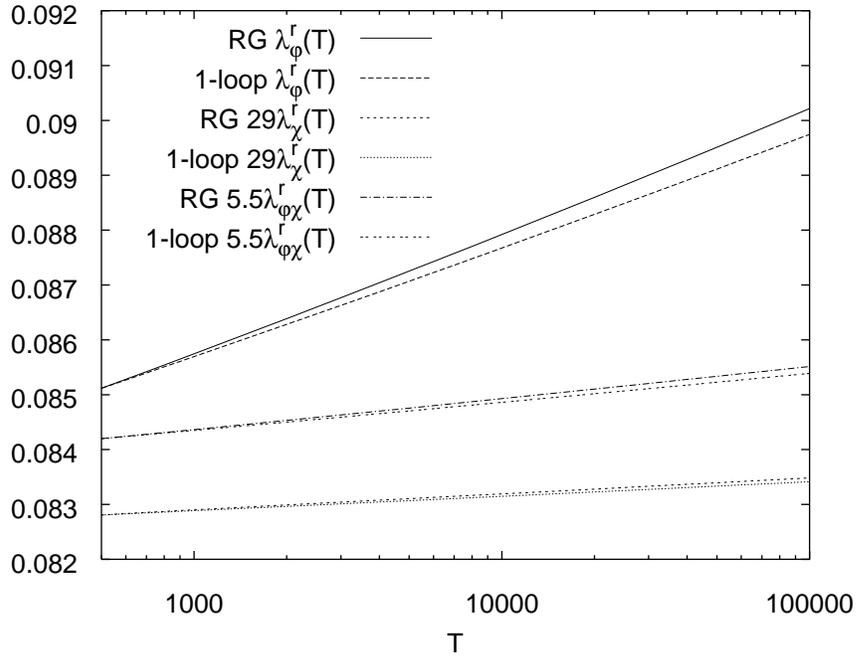}}  
\caption[The high temperature evolution of the couplings]
{The high temperature evolution of the couplings. ``RG''
refers to the Wilson RG approach of this paper while ``1-loop'' refers
to the numerical solution of equations
(\protect\ref{eq:beta1})--(\protect\ref{eq:beta3}). The initial conditions
for the one-loop integration are chosen so that the two approaches agree
at $T=500$. ($m^2_{r}=\mu^2_r=1$, $\lambda_{\varphi}^{r}=0.1$,
$\lambda_{\chi}^{r}=0.003$, $\lambda_{\varphi\chi}^{r}=0.016$)}
\label{fig:coupvbigt}
\end{figure} 
An example of the evolution of the couplings is shown in
Fig.\ \ref{fig:coupvbigt}, which also presents the running according
to the one-loop $\beta$-functions for comparison.  We see that at least
over the limited temperature range presented the two methods agree
reasonably well. Note that the renormalized couplings in
Fig.\ \ref{fig:coupvbigt} are chosen so that the theory is symmetric
at $T=0$ but broken at high temperature. The transition to the broken
phase occurs at $T_{c}\approx 50$.

We conclude this section by discussing an interesting phenomenon
related to the evolution of the couplings. Consider choosing
renormalized couplings which satisfy
\begin{equation}
\lambda_{\varphi}\gg \lambda_{\chi} \gtrsim \lambda_{\varphi\chi}\;.
\end{equation}
From (\ref{eq:beta1})--(\ref{eq:beta3}) it is clear that according to the
one-loop $\beta$-functions the couplings will flow in such a way that
eventually $\lambda_{\varphi\chi}(t) >3\lambda_{\chi}(t)$.  For
suitable initial values this occurs when all couplings are still
small, so that the loop expansion ought to be valid. This means that
the symmetry of the theory should be given by Eq.\ (\ref{eq:1loop}),
and we see that the running of the couplings can alter the symmetry of
the theory at high temperature. This effect can be exploited for model
building, in that it provides a ``natural'' way of having phase
transitions at very high temperature. For example, one can have
symmetry restoration set in above the GUT scale without having to fine
tune the couplings to one part in $10^{16}$. Similarly, one can
construct models that are broken at low temperature, get restored at
some $T_c \sim {\cal O}(m/\sqrt\lambda)$, and then get rebroken at a
much higher temperature. For a practical application of this kind of
effect to the strong $CP$ problem and the baryon asymmetry see
\cite{mohapat}.

The above discussion was based on the assumption that the couplings
evolve at high temperature according to their one-loop
$\beta$-functions. We have already stated that this is approximately
true, and we therefore expect that the evolution may indeed reverse
the inequality $3\lambda_{\chi}>\lambda_{\varphi\chi}$ at high
temperature. We wish to investigate the question if this really
leads to symmetry breaking, as predicted by (\ref{eq:1loop}).
Unfortunately the errors inherent in the numerical solution of our
evolution equations do not allow us to vary the temperature over the
15 or so orders of magnitude necessary to achieve a significant change
in the couplings. We therefore focus on the situation where one starts
with $\lambda_{\varphi\chi}$ less than, but very close to,
$3\lambda_{\chi}$, so that even a moderate change in $T$ can alter
their relationship. 

\begin{figure}[htb]
\epsfxsize=5in
\centerline{\epsfbox{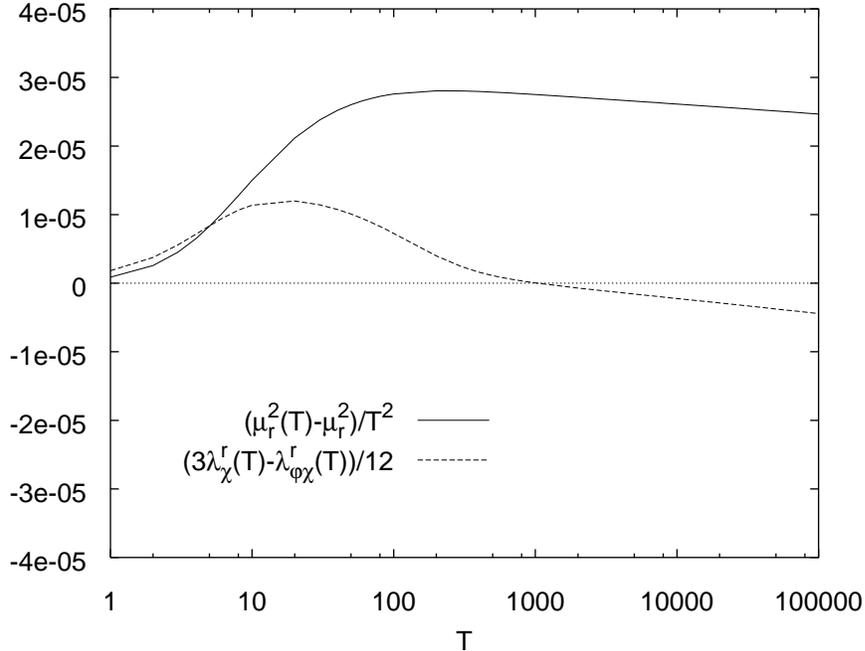}}  
\caption[The temperature dependence of two combinations of parameters
in the Wilson RG approach]
{The temperature dependence of two combinations of parameters
in the Wilson RG approach. $\mu^2_r(T)$ is rapidly increasing long
after the couplings satisfy the naive inequality which predicts
symmetry breaking. ($m^2_r=\mu^2_r=1,\;
\lambda^r_{\varphi}=0.1,\; \lambda^r_{\chi}=1\times 10^{-3},\;
\lambda^r_{\varphi\chi}=2.99\times10^{-3}$)}
\label{fig:coupbreak}
\end{figure}

Bearing in mind the above restrictions we choose
$\lambda^r_{\varphi}=0.1$, $\lambda^r_{\chi}=1\times 10^{-3}$, and
$\lambda^r_{\varphi\chi}=2.99\times10^{-3}$. If one numerically integrates
equations (\ref{eq:beta1})--(\ref{eq:beta3}) using these couplings as
initial conditions at $t=0$, the one-loop calculation (\ref{eq:1loop})
predicts a symmetry breaking phase transition at $T\approx 450$.  The
results of the Wilson RG integration are presented in Fig.\
\ref{fig:coupbreak}.
It is clear from the figure that the coefficient
of the $T^2$ term in $\mu^2_r(T)$ is not proportional to
$3\lambda^r_{\chi}-\lambda^r_{\varphi\chi}$ as predicted by Eq.\
(\ref{eq:1loop}). In addition we see that even long after the
perturbative coefficient has turned negative, $\mu^2_r(T)$ is still a
rapidly increasing function of temperature.  The phase transition
predicted by the one-loop result is absent. This is not to say, of
course, that a phase transition may not take place at much higher
temperature. After all $\mu^2_r(T)/T^2$ is decreasing, and if this
continues the symmetry will indeed by broken at very large $T$.  As
stated above, numerical errors prevent us from investigating this
region directly. 

Finally, we point out that for the initial conditions of Fig.\
\ref{fig:coupbreak}, $\lambda^r_{\chi}(T)$ does {\em not} run
according to its one-loop $\beta$-function over the temperature range
considered.  In fact, $\lambda^r_{\chi}(T)$ is decreasing very slowly
even at $T=10^5$. This shows that while generically the high
temperature behavior of the couplings is given by their one-loop
$\beta$-functions (Fig.\ \ref{fig:coupvbigt}), this need not be so.
The ``anomalous' behavior of $\lambda^r_{\chi}(T)$ for the case at
hand is due to the presence of two vastly different mass scales at
high temperature. When $m^2_r(T)$ and $\mu^2_r(T)$ differ by several
orders of magnitude, some terms in the flow equations ``decouple''
much earlier than others.  Under these circumstances our approach is
not expected to reproduce the one-loop result as the latter is blind to
the presence of mass scales.

\section{Conclusions}
\label{sec:concl}

In this work we investigated the high temperature phase structure of a
simple two scalar theory by solving approximately a non-perturbative
RG equation for the effective potential. The solution consisted of
assuming a polynomial expression for the potential and then
numerically integrating the resulting coupled flow equations for the
coefficients. Our main result was obtained in section~\ref{sec:z2z2}:
according to our non-perturbative method high temperature symmetry
breaking (or non-restoration) does exist. In addition, we found that
the phenomenon takes place roughly for those values of the couplings
that satisfy the inequalities obtained from perturbation theory. The
total volume of parameter space that yields high temperature symmetry
breaking is only slightly reduced compared to the one-loop prediction.

We also saw that the critical temperature obtained from perturbation
theory does not agree particularly well with our numerical results,
even for reasonably small couplings ($<0.1$).  The reason for this is
that the negative cross-coupling between the fields which drives the
symmetry breaking must by somewhat larger than the value predicted by
the loop expansion. This leads to a situation where the perturbative
estimate for the critical temperature becomes totally unreliable as
the cross coupling gets close to the boundary region. If one stays far
away from the boundary the perturbative and numerical results approach
each other to within 15\%, which is about what one would expect. We
also demonstrated that the critical temperature depends significantly
on the value of the quartic coupling of the unbroken field, in contrast
to what is predicted by the one-loop calculation.

In section~\ref{sec:couplings} we discussed the behavior of the coupling 
constants at high temperature. In the $\varphi^4$ case the quartic
coupling was shown to increase at very high temperature, but not before
being significantly reduced during an intermediate regime. For very
large T the evolution was shown to be approximately given by the one-loop
$\beta$-function of ordinary perturbation theory.

For the ${\Bbb Z}_2 \times{\Bbb Z}_2$ theory we focused on the effect of
the running couplings on the symmetry of the theory. It was shown that
generically the couplings evolve according to their one-loop
$\beta$-functions at high temperature. However, we also demonstrated
that it is {\em not} correct to draw conclusions about the symmetry of
the theory based on the perturbative formula for the thermal mass and
the running of the couplings. Naively this idea can be used to produce
``natural'' very high temperature phase transitions induced by the
evolution of the couplings. Our RG approach shows that this fails for
at least three reasons. First of all, even for generic initial
conditions, the couplings evolve according to their one-loop
$\beta$-functions only at very high $T$, and they are significantly
renormalized by the time they reach this region. Because the couplings
evolve very slowly after the initial decrease it is important to have
these ``initial values'' in order to start the high temperature
running. Using the zero temperature parameters would not suffice to
get an order of magnitude estimate of $T_c$ even if the rest of the
reasoning was correct. Second, any attempt to cause high temperature
symmetry breaking induced by the running of the couplings requires
that one of the quartic couplings be significantly larger than the
others. This introduces two widely different mass scales (the two
thermal masses), with the result that some of the couplings don't
evolve according to their one-loop $\beta$-functions even at very
large $T$. Lastly, and most importantly, we showed that even when the
couplings have evolved to fulfill the naive inequality that predicts
symmetry breaking, it does not happen.

The above remarks are best illustrated by the example discussed in
section \ref{sec:couplings}. For the renormalized parameters chosen
there the one-loop evolution of the couplings (Eqs.\
(\ref{eq:beta1})--(\ref{eq:beta3})) combined with the perturbative
formula for the thermal mass (Eq.\ (\ref{eq:1loop})) predicts a
symmetry breaking phase transition at $T\approx 450$. In contrast,
Fig.\ \ref{fig:coupbreak} shows that the relevant thermal mass is a
rapidly increasing function of temperature even at $T=10^5$. It is
possible that the symmetry will be broken at much higher $T$, but if
this happens it will be at a temperature many orders of magnitude
larger than that predicted by perturbation theory.

Finally we would like to comment on the generality of our
conclusions. All of our numerical results were obtained for the simple
${\Bbb Z}_2 \times{\Bbb Z}_2$ model. However, the differences between our
RG approach and the standard one-loop treatment arise because our method
correctly takes into account the decoupling of massive particles from the
theory as the infrared cutoff is lowered. This feature is intrinsic to the
method and independent of the particular model studied, which leads us to
expect similar deviations from the perturbative results for other models.

\section*{Acknowledgements}
I would like to thank Mark Alford and Brian Greene for many helpful
discussions. This work was supported in part by the National Science
Foundation.

\appendix
\section*{Appendix: Critical Behavior}

The main emphasis of this work does not rely on the details of the
phase {\em transitions} we have studied. Rather, we have been
interested in establishing the {\em existence} of symmetry breaking phase
transitions, and in showing that the broken symmetry state can persist
at arbitrarily high temperature. In this context it is only of
tangential interest as to what critical exponents characterize the
transition or even whether it is first or second order.  In fact, we
could have studied the persistence of the broken state by starting in
an asymmetric vacuum at $T=0$, in which case there would have been no
phase transition at all\footnote{So long as the couplings obey the
correct inequalities, see the discussion above
(\ref{eq:tcnaive}).}. Similarly, the evolution of parameters at very
high temperature does not depend on the detailed dynamics of a phase
transition that has taken place at much lower $T$.  The nature of the
phase transition is important, however, in understanding the behavior
of the renormalized parameters near $T_{c}$, shown in
Figs.~\ref{fig:paramvt}--\ref{fig:lambdasvt}. For this reason we
include here a brief discussion of the phase transitions studied in
sections \ref{sec:phi4} and \ref{sec:z2z2}.  Detailed applications of
Wilson type RG equations to critical phenomena can be found in
\cite{all,morristrunc,mark.wo.eps,wettphi4t,wett2scalt,wettcrit}.
 
We begin with the well known case of $\lambda\varphi^4$ theory. In order
to study the character of a transition it is helpful to cast the flow
equations into scale invariant form by using the proper dimensionless
couplings. To achieve this consider the theory at high temperature. Once
$\Lambda<2\pi T$ all non-zero Matsubara modes have been integrated out
and we are left with an effective three dimensional theory for the
zero mode (see (\ref{eq:dUhighT})). The coupling in this theory, which
has dimension one, is $\lambda_4 T$. The appropriate dimensionless mass
and coupling parameters of the effective theory are hence 
\begin{equation}
\kappa(\Lambda,T) = \frac{m^2(\Lambda,T)}{\Lambda^2}
\end{equation}
and
\begin{equation}
h(\Lambda,T) = \frac{\lambda_4(\Lambda,T) T}{\Lambda}\,.
\end{equation}
Rewriting equations (\ref{eq:dm}) and (\ref{eq:dl4}) in terms of
$\kappa$ and $h$ and using $t=\ln(\Lambda_0/\Lambda)$ we obtain the
desired form:
\begin{eqnarray}
\frac{d\kappa}{dt}&=&2\kappa + \frac{3 h}{2\pi^2(1+\kappa)}\, ,\label{eq:dk}
\\
\frac{d h}{dt}&=&h - \frac{9 h^2}{2\pi^2(1+\kappa)^2}\, \label{eq:dh}
\end{eqnarray}
(here we have truncated by setting $\lambda_n$=0 for $n \geq 6$, as
before).  Equations (\ref{eq:dk}) and (\ref{eq:dh}) have two fixed
points: the Gaussian (trivial) fixed point at
$\kappa_\star=h_\star=0$, and the Wilson fixed point (WFP) at
$\kappa_\star=-1/7$ and $h_\star=8\pi^2/49$. Linearizing around the
fixed points one finds that the former is completely unstable in the
IR ($t\rightarrow\infty$), while the latter is a saddle point. Hence
in order for a flow to end up at the WFP one needs to
fine-tune one linear combination of UV couplings, which in our case
amounts to choosing 
$T=T_{c}$. Consequently $\kappa\rightarrow\kappa_\star$ and
$h\rightarrow h_\star$ as $t\rightarrow \infty$ at the critical
temperature, which means that the transition is second order and that
\begin{eqnarray} 
m^2_r(T_{c}) = \lim_{\Lambda\rightarrow0} m^2(\Lambda,T_{c}) \sim
\lim_{\Lambda\rightarrow0}\Lambda^2\kappa_\star = 0 \, ,\label{eq:mc}
\\
\lambda_{4r}(T_{c})= \lim_{\Lambda\rightarrow0}
\lambda_4(\Lambda,T_{c}) \sim \lim_{\Lambda\rightarrow0}
\frac{\Lambda h_\star}{T_{c}}=0. \label{eq:lc}
\end{eqnarray}
This explains the behavior observed in
Figs.\ \ref{fig:rhovlam}--\ref{fig:paramvt} near $T_{c}$. 

We point out that the above conclusions are independent of the
parameterization used for the the flow equations, as they should
be. For example, consider starting with the flow equations in the
broken phase, (\ref{eq:drho}) and (\ref{eq:dl4b}). The appropriate
dimensionless parameters of the effective three dimensional theory are
now $\tilde{\kappa} = \rho_0(\Lambda,T)/\Lambda T$ (recall that the
field has dimension one-half in three dimensions) and $\tilde{h} =
\lambda_4(\Lambda,T) T/\Lambda$. In terms of these variables
(\ref{eq:drho}) and (\ref{eq:dl4b}) take a scale invariant form
similar to (\ref{eq:dk}) and (\ref{eq:dh}). One again finds two fixed
points, the Gaussian at $\tilde{\kappa}_\star=3/4\pi^2$,
$\tilde{h}_\star =0$ and the WFP at $\tilde{\kappa}_\star=1/4\pi^2$,
$\tilde{h}_\star = 2\pi^2$.  The former is again completely unstable
and the latter is again a saddle point, so our above conclusions
remain valid.

We now turn to the two scalar theory of section \ref{sec:z2z2}.
Just as above we consider the effective three dimensional high
temperature theory and rewrite the flow equations
(\ref{eq:dmz2z2})--(\ref{eq:dlpxz2z2}) in terms of the dimensionless
variables $\kappa_\varphi = m^2(\Lambda,T)/\Lambda^2$, $\kappa_\chi
= \mu^2(\Lambda,T)/\Lambda^2$, $h_\varphi =
\lambda_\varphi(\Lambda,T)T/\Lambda$, $h_\chi =
\lambda_\chi(\Lambda,T)T/\Lambda$, and $h_{\varphi\chi} =
\lambda_{\varphi\chi}(\Lambda,T)T/\Lambda$. This results in the
following set of scale invariant equations:
\begin{eqnarray}
\frac{d\kappa_\varphi}{dt} &=& 2\kappa_\varphi +
\frac{1}{2\pi^2}\left[\frac{3h_\varphi}{1+\kappa_\varphi}-
\frac{h_{\varphi\chi}}{1+\kappa_\chi}\right]\, , \\
\frac{d\kappa_\chi}{dt} &=& 2\kappa_\chi +
\frac{1}{2\pi^2}\left[\frac{3h_\chi}{1+\kappa_\chi}-
\frac{h_{\varphi\chi}}{1+\kappa_\varphi}\right]\, , \\
\frac{dh_\varphi}{dt}&=&
h_\varphi-\frac{1}{2\pi^2}\left[\frac{9h_\varphi^2}{(1
+\kappa_\varphi)^2}+\frac{h_{\varphi\chi}^2}{(1+\kappa_\chi)^2}\right]\,
, \\ \frac{dh_\chi}{dt}&=&
h_\chi-\frac{1}{2\pi^2}\left[\frac{9h_\chi^2}{(1
+\kappa_\chi)^2}+\frac{h_{\varphi\chi}^2}{(1+\kappa_\varphi)^2}\right]\,
, \\ \frac{dh_{\varphi\chi}}{dt} &=&
h_{\varphi\chi}-\frac{1}{2\pi^2}\left[ \frac{3h_\varphi
h_{\varphi\chi}}{(1+\kappa_\varphi)^2} + \frac{3h_\chi
h_{\varphi\chi}}{(1+\kappa_\chi)^2} 
-\frac{4h_{\varphi\chi}^2}{(1+\kappa_\varphi)(1+\kappa_\chi)}\right].\quad
\end{eqnarray}
At this point one could easily determine the fixed points of the above
system, investigate their stability, etc., but this is outside the
scope of the present article. Rather, we note that since the 
$\chi$-field remains massive during the transitions studied in
section~\ref{sec:z2z2} it decouples in
the IR, in the sense that $\kappa_\chi\rightarrow\infty$ like
$1/\Lambda^2$ as $\Lambda\rightarrow 0$. For small $\Lambda$ we are therefore
left with the reduced system 
\begin{eqnarray}
\frac{d\kappa_\varphi}{dt} &=& 2\kappa_\varphi +
\frac{1}{2\pi^2}\frac{3h_\varphi}{1+\kappa_\varphi}\, , \label{eq:mcz2z2}
\\
\frac{dh_\varphi}{dt}&=& h_\varphi-\frac{1}{2\pi^2}\frac{9h_\varphi^2}{(1
+\kappa_\varphi)^2}\, ,\label{eq:hphiz2z2}
\\
\frac{dh_\chi}{dt}&=& h_\chi-\frac{1}{2\pi^2}
\frac{h_{\varphi\chi}^2}{(1+\kappa_\varphi)^2} \, , \label{eq:hpcz2z2}
\\
\frac{dh_{\varphi\chi}}{dt} &=& h_{\varphi\chi}-\frac{1}{2\pi^2}
\frac{3h_\varphi h_{\varphi\chi}}{(1+\kappa_\varphi)^2}\, .\label{eq:lcz2z2}
\end{eqnarray}
This system has the same two fixed points we found in the
$\lambda\varphi^4$ case above, namely the trivial one and the WFP at
$\kappa_\varphi=-1/7$, $h_\varphi=8\pi^2/49$,
$h_\chi=h_{\varphi\chi}=0$. What has changed is the stability of these
points. While the Gaussian fixed point is still completely unstable,
the WFP now has three unstable directions instead of just one. Thus it
seems that fine-tuning {\em one} linear combination of UV couplings
(by adjusting $T$) is not sufficient for a flow to end up at the WFP.
This, combined with the absence of other sufficiently stable fixed
points, seems to indicate that the system does not undergo second
order phase transitions.

Given the above analysis one may well ask why the order parameter and
the couplings seem to vanish continuously near the critical
temperature, as shown in Figs.~\ref{fig:massesvt} and
\ref{fig:lambdasvt}. The reason is that $\kappa_\varphi$ and
$h_{\varphi}$ flow to the WFP despite the instability in the other two
directions in coupling space. This happens because the
$\varphi$-sector of the theory completely decouples from the
$\chi$-sector (but not vice versa), as can be seen from
(\ref{eq:mcz2z2})--(\ref{eq:lcz2z2}). In fact, (\ref{eq:mcz2z2}) and
(\ref{eq:hphiz2z2}) are exactly the same as the equations for the
simple $\varphi^4$ case, (\ref{eq:dk}) and (\ref{eq:dh}). Consequently
all the reasoning for that case goes through and it suffices to tune
one parameter (the temperature) in order to flow to the fixed
point\footnote{It is actually not quite that simple. The fact that the
WFP is a saddle point in the $\varphi$ subspace indicates that one
needs to fine-tune only one linear combination of couplings to flow to
it. The important question is then for what values of the parameters
adjusting the temperature enables one to achieve a correct linear
combination. This is the question we have studied in detail in
section~\ref{sec:z2z2}, with the conclusion that the allowed parameter
values are roughly those predicted by perturbation theory. In this
appendix we argue only that {\it if} it is possible to produce a
symmetry breaking phase transition by varying $T$, then this
transition is second order and the critical behavior is equivalent to
that of the ${\Bbb Z}_2$ model.}. For our quartic truncation the
transition is thus predicted to be second order and $m^2_r(T_c)$ and
$\lambda_{\varphi}^r(T_c)$ flow as in (\ref{eq:mc}) and (\ref{eq:lc})
as $\Lambda\rightarrow 0$. The behavior of $h_{\varphi\chi}$ and
$h_\chi$ at the phase transition can be found by plugging the fixed
point values $\kappa_\varphi = -1/7$ and $h_{\varphi}=8\pi^2/49$ into
(\ref{eq:hpcz2z2}) and (\ref{eq:lcz2z2}). These equations are then
easily integrated, with the result $h_{\varphi\chi}\sim \exp(2t/3)$
and $h_\chi\sim c_1\exp (t) - c_2\exp (4t/3)$, where $c_1$ and $c_2$
are positive constants. For $\Lambda\rightarrow 0$ we thus obtain
$\lambda_{\varphi\chi}(\Lambda,T_{c}) \sim \Lambda^{1/3}$ and
$\lambda_{\chi}(\Lambda,T_{c}) \sim c_1-c_2/\Lambda^{1/3}$. This
behavior is observed in Fig.~\ref{fig:lambdasvt}. The fact that
$\lambda_{\chi}$ does not go negative in the figure is simply due to
the temperature resolution used. For $T$ very close to
$T_{c}$\footnote{For the parameters used in the figure this requires
$|T-T_{c}|/T_{c} \sim 10^{-7}$.} one indeed finds
$\lambda_{\chi}^r(T)<0$.

We conclude with a few comments. First, we have verified numerically
that close to $T_{c}$ the RG trajectories of the full theory do get
attracted to the WFP in the $\varphi$ subspace. This shows that the
WFP is the relevant one for our phase transitions. Second, we point
out that the decoupling of the $\varphi$-sector from the $\chi$-sector
at the phase transition will occur even if the evolution equations are
not truncated at quartic order in the fields as was done above. The
reason for this is simple: any diagram that contributes to the running
of some coupling $\lambda_{\varphi}^{(n)}$ (corresponding to an
operator $\varphi^n$) is either built from vertices that have no
$\chi$ legs or else necessarily contains closed $\chi$-loops. The
former are the same vertices present in $\varphi^4$ theory, and the
latter will be highly suppressed in the IR when $\kappa_\chi =
\mu^2/\Lambda^2 \rightarrow \infty$. The critical behavior of our two
scalar theory near the inverse phase transitions under consideration
is thus equivalent to that of the ${\Bbb Z}_2$ model for any
polynomial truncation of the effective potential. Finally, a comment
regarding the behavior of $\lambda_{\chi}$ at $T_{c}$. While it may
seem odd that this coupling goes to minus infinity as the cutoff is
lowered, this does not mean that the theory is unbounded. It must be
remembered that without our truncation higher order terms would be
present and that the four dimensional couplings do not have a simple
physical interpretation in the critical theory. To illustrate this
point consider the ${\Bbb Z}_2$ model, and the coupling $\lambda_8$
corresponding to the operator $\varphi^8$. At high temperature the
dimensionless effective three dimensional coupling relevant for the
investigation of fixed points is $h_8 = T^3\Lambda\lambda_8$. The
fixed point value of $h_{8}$ turns out to be negative
\cite{mark.wo.eps}, which means that
$\lambda_8(\Lambda,T_{c})\rightarrow -\infty$ as $\Lambda\rightarrow
0$. In fact it is clear that all $\lambda_n$ with $n\geq 8$ will
diverge at $T_{c}$. The point is that the physics near the phase
transition is parameterized by the critical exponents of the effective
three dimensional theory and not by the four dimensional couplings.

\end{document}